\documentclass[conf]{new-aiaa}
\usepackage[utf8]{inputenc}

\usepackage{graphicx}
\usepackage{subcaption}
\usepackage{siunitx}
\usepackage{amsmath,amsfonts,amssymb,amsthm}
\usepackage{nicefrac}
\usepackage[version=4]{mhchem}
\usepackage{siunitx}
\usepackage{longtable,tabularx}
\usepackage{mathtools}
\usepackage{commath}
\usepackage{booktabs}
\usepackage[table,xcdraw]{xcolor}
\usepackage[skip=0\baselineskip]{caption}
\def\multiset#1#2{\ensuremath{\left(\kern-.3em\left(\genfrac{}{}{0pt}{}{#1}{#2}\right)\kern-.3em\right)}}

\newtheorem{theorem}{Theorem}

\newtheorem*{myproof*}{Proof}

\newcommand{\ra}[1]{\renewcommand{\arraystretch}{#1}}

\title{Nonlinear Trajectory-Based Region of Attraction Estimation for Aircraft Dynamics Analysis}

\author{Brian Lai\footnote{Undergraduate Research Assistant, Department of Mechanical \& Aerospace Engineering, 98 Brett Road, Piscataway, NJ 08854}}
\affil{Rutgers University, Piscataway, New Jersey, 08854}
\author{Torbjørn Cunis\footnote{Research Fellow, Department of Aerospace Engineering, 1320 Beal Avenue, Ann Arbor, MI 48109-2140, AIAA Member}}
\affil{University of Michigan, Ann Arbor, Michigan, 48109}
\author{and}
\affil{\vspace{-12pt} \quad}
\author{Laurent Burlion\footnote{Assistant Professor, Department of Mechanical \& Aerospace Engineering, 98 Brett Road, Piscataway, NJ 08854, AIAA Member}}
\affil{Rutgers University, Piscataway, New Jersey, 08854}

\begin{document}

\maketitle

\begin{abstract}
Current flight control validation is heavily based on linear analysis and high fidelity, nonlinear simulations. Continuing developments of nonlinear analysis tools for flight control has greatly enhanced the validation process. Many analysis tools are reliant on assuming the analytical flight dynamics but this paper proposes an approach using only simulation data. First, this paper presents improvements to a method for estimating the region of attraction (ROA) of nonlinear systems governed by ordinary differential equations (ODEs) based only on trajectory measurements. Faster and more accurate convergence to the true ROA results. These improvements make the proposed algorithm feasible in higher-dimensional and more complex systems. Next, these tools are used to analyze the four-state longitudinal dynamics of NASA's Generic Transport Model (GTM) aircraft. A piecewise polynomial model of the GTM is used to simulate trajectories and the developed analysis tools are used to estimate the ROA around a trim condition based only on this trajectory data. Finally, the algorithm presented is extended to estimate the ROA of finitely many equilibrium point systems and of general equilibrium set (arbitrary equilibrium points and limit cycles) systems.   
\end{abstract}

\section{Introduction}

Extensive validation of flight control systems is needed to ensure safety in service. It is common practice to analyze flight dynamics using linear analysis tool around many trim conditions. Additional Monte Carlo simulations of the nonlinear flight dynamics may be used to supplement missing nonlinear phenomenon \cite{Chakraborty_GTM}. However, flight dynamics beyond stall are highly nonlinear and often unstable, resulting in numerous nonlinear effects. \cite{Goman}. As a result, nonlinear effects can go unnoticed by Monte Carlo simulations. A well-studied example is the loss of several F/A-18 aircraft due to a nonlinear loss-of-control phenomenon now known as the falling leaf mode \cite{Chakraborty_Falling_Leaf,Chakraborty_Falling_Leaf_nonlinear}. Much work has been done to analyze and design control systems for such nonlinear phenomenon \cite{Chakraborty_GTM,Cunis2019,GTM_ex1,Chakraborty_Falling_Leaf_nonlinear}.
\par
These works often utilize Lyapunov-based region of attraction (ROA) estimation techniques for nonlinear systems governed by ordinary differential equations (ODEs). Numerous advancements in Lyapunov-based methods have made such analysis possible, including works on robustness of estimation \cite{Topcu1}, uncertain nonlinear systems \cite{Topcu2} and composite Lyapunov functions \cite{Tan}. Many approaches make use of polynomial sum-of-squares optimization, which can only be applied to polynomial vector field dynamics \cite{Parrilo}. Some approaches however have shown successful ROA estimation using only simulation data \cite{Colbert, Topcu2008}. Such approaches do not require an analytical representation of dynamics but only simulation data from a finite number of initial conditions. 
\par
The objective of this paper is to use a trajectory-based Lyapunov ROA estimation method to analyze the flight dynamics of a scale aircraft. It is of interest to use simulation-based estimation methods for flight control verification as aircraft dynamics models make assumptions to the true dynamics and can have limited range of fidelity \cite{GTM_piecewise}. Although an aircraft dynamics model is used in this study to generate trajectory data, this work serves as a proof of concept for ROA estimation where sufficient flight trajectory data has been collected, but an analytical representation of the aircraft dynamics remains unknown. 


\section{NASA GTM Piecewise Polynomial Modeling}

This work will utilize NASA's Generic Transport Model (GTM), which represents a \SI{5.5}{\percent} scale aircraft, and provides extensive full-envelope aerodynamic data from wind tunnel studies. \cite{NASA_GTM}. The open source 6 degree-of-freedom Simulink model of the GTM has been an important resource for modeling, analysis, and control design in the aerospace community. 

This study considers the longitudinal dynamics of the GTM. Relevant nomenclature and parameter values are given in tables \ref{GTM Variables} and \ref{GTM Constants}. The longitudinal dynamics consist of 4D state $\mathcal{X} = [V_\mathrm{A},\gamma_\mathrm{A},q,\alpha]^\mathrm{T}$, and control input $U = [\eta,F]^\mathrm{T}$. The four state longitudinal dynamics of the GTM are described by the equations of motion (\ref{Longitudinal EOM}).

\begin{table}[ht] \centering
\ra{1.3}
\caption{NASA GTM Variable Parameters}
\label{GTM Variables}
\begin{tabular}{@{}lll@{}}
\toprule[1pt]\midrule[0.3pt]
$\alpha$ & Angle of Attack & \SI{}{\radian}\\
$q$ & Pitch Rate & \SI{}{\radian\per\second}\\
$\hat{q}$ & Normalized Pitch Rate $\left( \hat{q} = \frac{\bar{c} q}{2V_\mathrm{A}} \right)$ & \SI{}{\radian} \\
$\gamma_\mathrm{A}$ & Air-Path Inclination Angle & \SI{}{\radian}\\
$V_\mathrm{A}$ & Air Speed Relative to Air & \SI{}{\meter\per\second}\\
$\eta$ & Elevator Deflection (negative if leading to positive pitch moment) & \SI{}{\radian} \\
$F$ & Thrust Force (positive along body x\textsubscript{f}-axis) & \SI{}{\N} \\
\midrule
$C_\mathrm{m}$ & Aerodynamic Moment Coefficient, Body y\textsubscript{f}-Axis & \SI{}{1} \\
$C_\mathrm{D}$ & Aerodynamic Drag Coefficient, Negative Air-Path x\textsubscript{a}-axis & \SI{}{1}  \\
$C_\mathrm{L}$ & Aerodynamic Lift Coefficient, Negative Air-Path z\textsubscript{a}-axis & \SI{}{1}  \\
$C_\mathrm{X}$ & Aerodynamic Force Coefficient, Body x\textsubscript{f}-Axis & \SI{}{1}  \\
$C_\mathrm{Z}$ & Aerodynamic Moment Coefficient, Body z\textsubscript{f}-Axis & \SI{}{1}  \\
\midrule[0.3pt]\bottomrule[1pt]
\end{tabular}
\end{table}

\vspace{5pt}

\begin{table}[ht]\centering
\ra{1.3}
\caption{NASA GTM Constant Parameters}
\label{GTM Constants}
\begin{tabular}{@{}lll@{}}
\toprule[1pt]\midrule[0.3pt]
$S$ & Wing Area & 0.550 m\textsuperscript{2} \\
$\bar{c}$ & Mean Aerodynamic Chord & \SI{0.280}{\meter} \\
$m$ & Aircraft Mass & \SI{26.190}{\kilogram} \\
$I_\mathrm{yy}$ & Body y\textsubscript{f}-Axis Moment of Inertia & \SI{5.768}{\kilogram\meter\squared} \\
$l_\mathrm{t}$ & Engine Vertical Displacement (positive along z\textsubscript{f}-axis) & \SI{0.100}{\meter} \\
$x_\mathrm{cg}, \ z_\mathrm{cg}$ & Longitudinal Position Center of Gravity & \SI{-1.450}{\meter}, \SI{-0.300}{\meter} \\
$x_\mathrm{cg}^\mathrm{ref}, \ z_\mathrm{cg}^\mathrm{ref}$ & Longitudinal Position Reference Center of Gravity & \SI{-1.460}{\meter}, \SI{-0.290}{\meter} \\
$\rho$ & Air Density & \SI{1.200}{\kilogram\per\meter\cubed}\\
$g$ & Gravitational Acceleration & \SI{9.810}{\meter\per\second\squared} \\
\midrule[0.3pt]\bottomrule[1pt]
\end{tabular}
\end{table}

\begingroup\makeatletter\def\f@size{9.5}\check@mathfonts
\def\maketag@@@#1{\hbox{\m@th\normalsize\normalfont#1}}%
\begin{subequations}
\begin{align}
\dot{V}_\mathrm{A} & = \ \frac{1}{m}\left(F\cos{\alpha} - \frac{1}{2}\rho SV_\mathrm{A}^2C_\mathrm{D}(\alpha,\eta,\hat{q})-mg\sin{\gamma_\mathrm{A}}\right) \\
\dot{\gamma}_\mathrm{A} & = \ \frac{1}{mV_\mathrm{A}}\left(F\sin{\alpha} + \frac{1}{2}\rho SV_\mathrm{A}^2C_\mathrm{L}(\alpha,\eta,\hat{q})-mg\cos{\gamma_\mathrm{A}}\right) \\
  \begin{split}
    \dot{q} & = \frac{1}{I_\mathrm{yy}}\left(l_\mathrm{t} F + \frac{1}{2}\rho S\bar{c}V_\mathrm{A}^2C_\mathrm{m}(\alpha,\eta,\hat{q}) - \frac{1}{2}\rho SV_\mathrm{A}^2C_\mathrm{Z}(\alpha,\eta,\hat{q}) \left(x_{\mathrm{cg}}^\mathrm{ref} - x_\mathrm{cg}\right) + \frac{1}{2}\rho SV_\mathrm{A}^2C_\mathrm{X}(\alpha,\eta,\hat{q})\left(z_\mathrm{cg}^\mathrm{ref} - z_\mathrm{cg}\right) \right)
  \end{split}
  \\
  \dot{\alpha} & = q - \dot{\gamma}
\end{align}
\label{Longitudinal EOM}
\end{subequations}
\endgroup

To analyze the longitudinal dynamics, a piecewise polynomial model of the aerodynamic coefficients developed by Cunis et al. is used \cite{Cunis2019,GTM_piecewise}. Polynomial models of aerodynamic coefficients have been integral in flight analysis for their continuity and infinite differentiability. Choosing a piecewise polynomial model allows for good aerodynamic coefficient representation throughout the entire flight envelope \cite{Cunis2019}. This model generates the trajectory data required for ROA estimation.


\section{Lyapunov Based ROA Estimation from Trajectory Data}

\subsection{Lyapunov Theory}
Consider an autonomous system given by the initial value problem
\begin{equation}
\label{ODE}
\dot{x}(t) = f(x(t)), \quad x(0) = x_0
\end{equation}
\noindent with $f:\mathbb{R}^n \rightarrow \mathbb{R}^n$ and $x_0 \in \mathbb{R}^n$; and assume that a solution map $g(x,t)$ exists such that $\frac{\mathrm{d}}{\mathrm{d}t} g(x,\tau) = f(g(x,\tau))$ for all $\tau \in \left[0, t\right]$ and $g(x,0) = x$. Furthermore, assume that $f(0) = 0$, thus implying that $x=0$ is an equilibrium point of the ODE.
The algorithm proposed in \cite{Colbert} estimates the largest set on which all points in the set are asymptotically stable to $x=0$. This set $S$ is denoted the region of attraction (ROA), and defined as 
\begin{equation}
\label{ROA 1 point}
  S := \{x \in \mathbb{R}^n : \lim_{t \to \infty}g(x,t) = 0  \}.  
\end{equation}

where $\lim_{t \to \infty}g(x,t) \coloneqq \infty$ for divergent trajectories. To estimate $S$, Lyapunov functions are used. Lyapunov functions are continuously differentiable functions $V:X \to \mathbb{R}$ on a compact set $X\subset \mathbb{R}^n$ such that 
\begin{subequations}
\begin{align}
    V(0) = 0 &\\
    V(x) > 0 & \ \textnormal{for} \ x \in X, \ x \neq 0 \\
    \dot{V}(x) = \nabla V(x)^Tf(x) < 0 & \ \textnormal{for} \ x \in X, \ x \neq 0
\end{align}
\label{Lyapunov Theorem}
\end{subequations}
 
\noindent Then, the ODE (\ref{ODE}) is asymptotically stable on any sublevel set $V_\gamma \subset X$ where 
 \begin{equation*}
     V_\gamma = \{x \in X : \ V(x) \leq \gamma\}.
 \end{equation*}
 Finally, considering a Lyapunov function of the form 
 \begin{equation}
     V(x) = \int_{0}^{\infty}  \norm{g(x,t)}^2 dt
     \label{Lyapunov Function}
 \end{equation}
 results in the important property that $ S = \lim_{\gamma \to \infty} V_\gamma $ for $V$ defined in (\ref{Lyapunov Function}) \cite{Colbert}.

\subsection{ROA Estimation Algorithm}
\indent The nonlinear ODE analysis techniques used are largely an extension of a paper by Colbert et. al \cite{Colbert} which presents a method for estimating the region of attraction of a nonlinear ODE system using only a finite number of known trajectories. The work of Colbert et al. \cite{Colbert} showed that this classical result can be used to determine the region of attraction $S$ when the solution map $g(x,t)$ is not known. In particular, a sum of squares (SOS) algorithm is used to compute $S$ based on trajectory data from several initial conditions (ICs) of the closed-loop system. 

The convergent trajectory data is notated as $a(i,j) = g(x_i,j\Delta t)$, for $j = 1,...,K$, where $\Delta t$ is the time step and $x_i$ is the convergent initial condition, and make the approximation \cite{Colbert} 
\begin{equation}
\label{approximation}
\begin{split}
    V(x_i) & = \int_{0}^{K\Delta t}\norm{g(x,t)}^2 dt + V(a(i,K)) \\ & \approx \int_{0}^{K\Delta t}\norm{g(x,t)}^2 dt \approx \sum_{j=0}^{K}\norm{a(i,j)}^2\Delta t  
\end{split}
\end{equation}
Next, the set of convergent initial conditions $\{x_i \ : \ i = 1,...,m\}$ is mapped to the set of outputs $\{y_i = \log_{10}{(1+V(x_i))} \ : \ i = 1,...,m\}$ via a sum of squares (SOS) polynomial $p(x)$. That is, 
\begin{equation*}
    p(x) = \sum\limits_{i=1}^{n}p_i(x)^2 \ , \ p_i(x) \in \mathbb{R}[x] \ , \ n \in \mathbb{N}
\end{equation*}
where $\mathbb{R}[x]$ is the set of multivariate polynomials of real coefficients of $x$. Further, $p(x)$ of degree $2d$ is SOS if and only if there exists a positive, semi-definite matrix $P\in \mathbb{R}^{q \times q}$ where $q = \binom{n+d}{n}$ such that
\begin{equation}
\label{SOS Ploynomial}
p(x) = Z_d(x)^\mathrm{T} P Z_d(x)
\end{equation}
where $Z_d(x) \in \mathbb{R}^q$ are the monomials of $x$ of degree $d$ or less. Then, an optimal such $p(x)$ is found, i.e. an optimal semi-definite $P\in \mathbb{R}^{q\times q}$, via the following semi-definite programming (SDP) optimization problem: \cite{Colbert}
\begin{equation}
\label{Semi Definite Prog Problem}
\begin{split}
    \displaystyle\min_{P\in \mathbb{R}^{q\times q}, \ \gamma \in \mathbb{R}^m} \ & -\sum\limits_{i=1}^{m} \gamma_i \\
    \textnormal{s.t.} \quad & Z_d(x_i)^\mathrm{T} P Z_d(x_i) - y_i \geq \gamma_i \ \\
    & y_i - Z_d(x_i)^\mathrm{T} P Z_d(x_i) \geq \gamma_i \ ; \ P  \succeq 0.
\end{split}
\end{equation}
The semi-definite optimization software, SeDuMi, is used to find an optimal matrix $P$, denoted $P^*$ \cite{sedumi, Sturm1999AMT}. Finally, the Lyapunov function of form (\ref{Lyapunov Function}) may be estimated by 
\begin{equation}
\label{Estimated Lyapunov Function}
    V^*(x) = 10^{Z_d(x)^\mathrm{T} P^* Z_d(x)} - 1 
\end{equation}
Denoting $\gamma^* = \displaystyle\max_{i , \in , 1,...,m}\{V^*(x_i)\}$, the estimated region of attraction $V_{\gamma^*}$ is expressed as
\begin{equation}
\label{Estimated Region of Attraction}
V_{\gamma^*} = {\{x \ : \ V^*(x)\leq \gamma^*\}}
\end{equation}

\subsection{Proposed Algorithm Improvements}
Two contributions are proposed to the original algorithm \cite{Colbert} which improve speed of convergence and accuracy of the ROA estimation. These contributions make the algorithm more feasible in more complex system and systems of dimension higher than two. Such improvements are critical in analyzing flight simulations more complex than two-state longitudinal short period motion.

\subsubsection{Trajectory Partitioning}

The algorithm presented in \cite{Colbert} is computationally intensive to converge to the real ROA because of the need to simulate many convergent trajectories. In order to reduce computation time, each trajectory is partitioned into multiple trajectories. For some initial condition $x_i$, take a data point $x_{i_1} = g(x_i,M\Delta t)$ on its trajectory as an additional initial condition. Notice that $x_{i_1}$ follows the same trajectory as $x_i$ onward, so map $x_{i_1}$ to $V\big(a(i,j)\big) = \sum_{j=M}^{K}\norm{a(i,j)}^2\Delta t $ in the SDP optimization problem (\ref{Semi Definite Prog Problem}). The process may also be extended to multiple partitions $x_{i_j}$ for each initial condition $x_i$. 
\par
\begin{figure}[ht]
       \centering
		\includegraphics[width=0.7\textwidth]{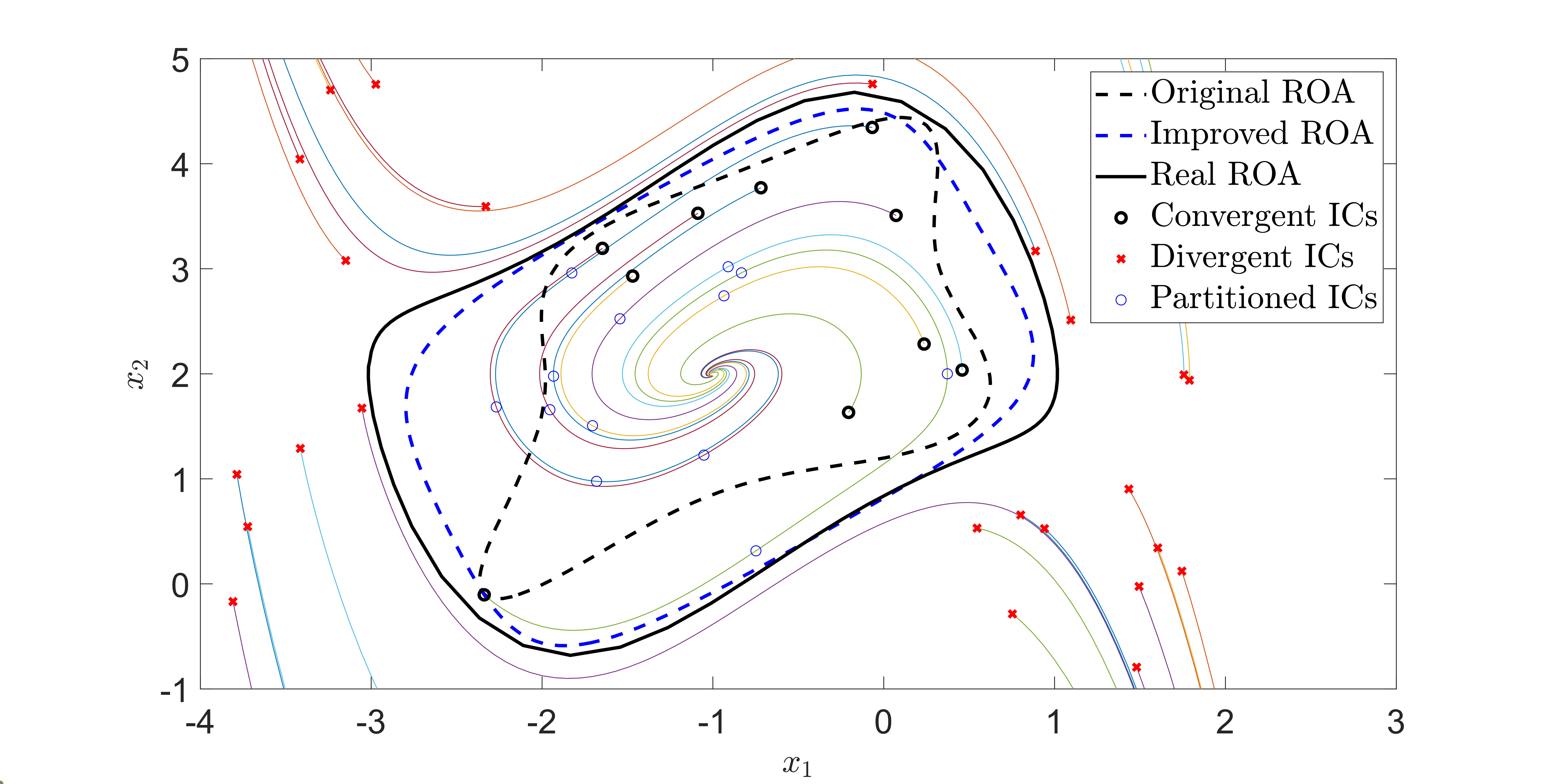}
		\caption{Estimated ROA of Van Der Pol Oscillator without and with partitioned ICs (10 and 23 ICs respectively). ``Original ROA" refers to the ROA calculated via the techniques of \cite{Colbert}. \textit{Improved ROA} refers to the ROA calculated with the addition of partitioned ICs.}  
		\label{trajectory partitioning}               
\end{figure}

Simulations using trajectory partitioning are done on the Van der Pol oscillator in reverse time defined as 
\begin{equation}
\begin{split}
    \dot{x}_1 & = -(x_2 - 2) \\
    \dot{x}_2 & = (x_1+1) + (x_2-2)((x_1+1)^2 - 1)
\end{split}
\end{equation}
Note that the system is center around equilibrium point $(-1,2) \in \mathbb{R}^2$ to show feasibility of a linear shift. Figure \ref{trajectory partitioning} shows an improved ROA estimation by adding initial conditions from partitioned trajectories.

Trajectory partitioning is useful in situations where only a limited number of trajectories are known. However, this algorithm creates a bias of initial conditions near the equilibrium point. As such, it is preferred to not use trajectory partitioning if a sufficient amount of trajectory data is available. 

\subsubsection{Guaranteeing Divergent ICs are Outside the ROA}

The algorithm presented in \cite{Colbert} utilizes only convergent trajectories to estimate the ROA. It is important however to test a combination of convergent and divergent initial conditions to find a convergent initial condition close to the boundary of the real region of attraction. An IC close to the boundary of the real ROA results in a large value of $\gamma^*$, thus a large region of attraction defined by (\ref{Estimated Region of Attraction}). 
\par
This algorithm guarantees that all convergent ICs tested are within the estimated ROA, but divergent ICs tested may still fall withing the estimated ROA. To guarantee that divergent ICs are outside the estimated ROA, the SDP problem (\ref{Semi Definite Prog Problem}) is modified by adding an additional constraint shown in (\ref{Semi Definite Prog Problem 2}).The set of divergent initial conditions of (\ref{ODE}) is notated as $\{x_{d_i} \ | \ i=1,...,s \}$. The SDP optimization problem is modified to
\begin{equation}
\label{Semi Definite Prog Problem 2}
\begin{split}
    \displaystyle\min_{P\in \mathbb{R}^{q\times q}, \ \gamma_i \in \mathbb{R}^m} \ & -\sum\limits_{i=1}^{m} \gamma_i 
    \\
    \textnormal{s.t.} \quad Z_d(x_i)^\mathrm{T} P Z_d(x_i) - y_i & \geq \gamma_i \ 
    \\
    y_i - Z_d(x_i)^\mathrm{T} P Z_d(x_i) & \geq \gamma_i \ ; \ P  \succeq 0. 
    \\
    \min\{Z_d(x_{d_i})^\mathrm{T} P Z_d(x_{d_i})\} & > \max\{Z_d(x_i)^\mathrm{T} P Z_d(x_i)\}
\end{split}
\end{equation}

By imposing that $\min\{Z_d(x_{d_i})^\mathrm{T} P Z_d(x_{d_i})\} > \max\{Z_d(x_i)^\mathrm{T} P Z_d(x_i)\}$, it is guaranteed that $V^*(x_{d_i}) > \gamma^*$, for all $x_{d_i}$, where $\gamma^* \coloneqq \displaystyle\max_{i , \in , 1,...,m}\{V^*(x_i)\}$. That is, that every divergent initial condition is outside ROA estimate $V_{\gamma^*}$. \\


\section{NASA GTM Aircraft Stability Analysis}

\subsection{ROA Estimation}
Trajectory data of the NASA GTM longitudinal dynamics (\ref{Longitudinal EOM}) were found using the piecewise polynomial model \cite{GTM_piecewise}. The longitudinal dynamics consist of a four-state $\mathcal{X} = [V_\mathrm{A},\gamma_\mathrm{A},q,\alpha]^\mathrm{T}$, and control input $U = [\eta,F]^\mathrm{T}$. 8565 trajectories were found under constant control input $\eta = \SI{0}{\degree}, F = \SI{20}{\N}$.

\begin{figure}[ht]
       \centering
		\includegraphics[width=0.95\textwidth]{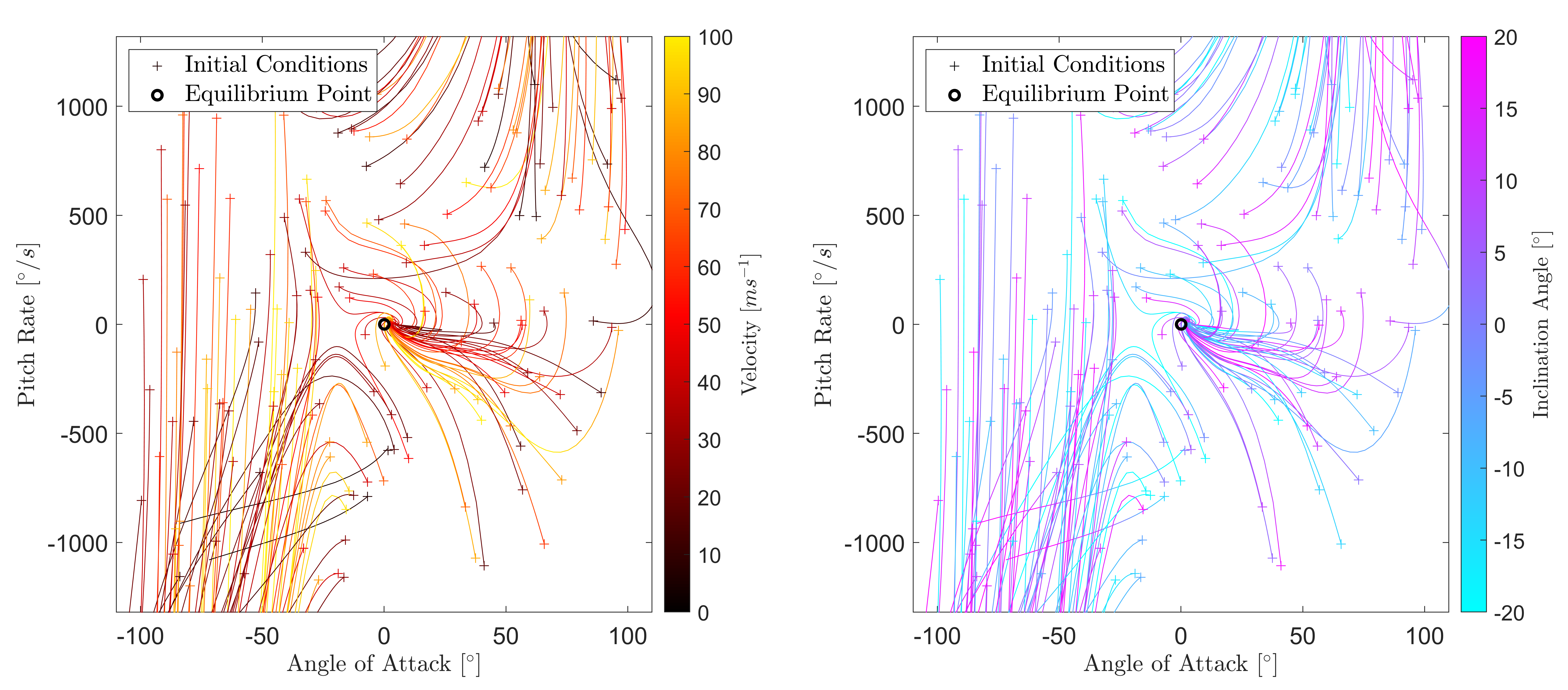}
		\caption{NASA GTM longitudinal dynamics simulations with fixed control: $\eta = \SI{0}{\degree}, F = \SI{20}{\N}$. Plot of 2\% of the 8565 trajectories used in ROA estimate. Left: velocity color scale. Right: inclination angle color scale.  }  
		\label{GTMTrajectories}               
\end{figure}

The run-time for each trajectory is 100 seconds. Trajectories are categorized as divergent if a state variable exceeds a set of limits on that variable. Otherwise, the trajectory is convergent. This classification is sufficient as the system contains a single asymptotically stable equilibrium. Initial conditions for trajectories are randomly chosen between bounds for each state variable. Table \ref{simulation bounds and limits} details these bounds and limits. 

\begin{table}[ht]\centering
\ra{1.3}
\begin{tabular}{@{}lrrcrr@{}}\toprule
& \multicolumn{2}{c}{Initial Condition Bounds} & \phantom{abc}& \multicolumn{2}{c}{Trajectory Limits}\\
\cmidrule{2-3} \cmidrule{5-6} 
& min & max && min & max\\ \midrule
$\alpha \ [\SI{}{\degree}]$ & -100 & 100 && -180 & 180 \\
$q \ [\SI{}{\degree\per\second}]$ & -1200 & 1200 && -2500 & 2500 \\
$\gamma_\mathrm{A} \ [\SI{}{\degree}]$ & -20& 20&& -45& 45\\
$V_\mathrm{A} \ [\SI{}{\meter\per\second}]$ & 5& 100&& 0& 300\\
\bottomrule
\end{tabular}
\caption{Initial condition bounds and trajectory limits for determining divergence in the NASA GTM longitudinal dynamics simulation.}
\label{simulation bounds and limits}
\end{table}

\begin{figure}[ht]
       \centering
		\includegraphics[width=0.68\textwidth]{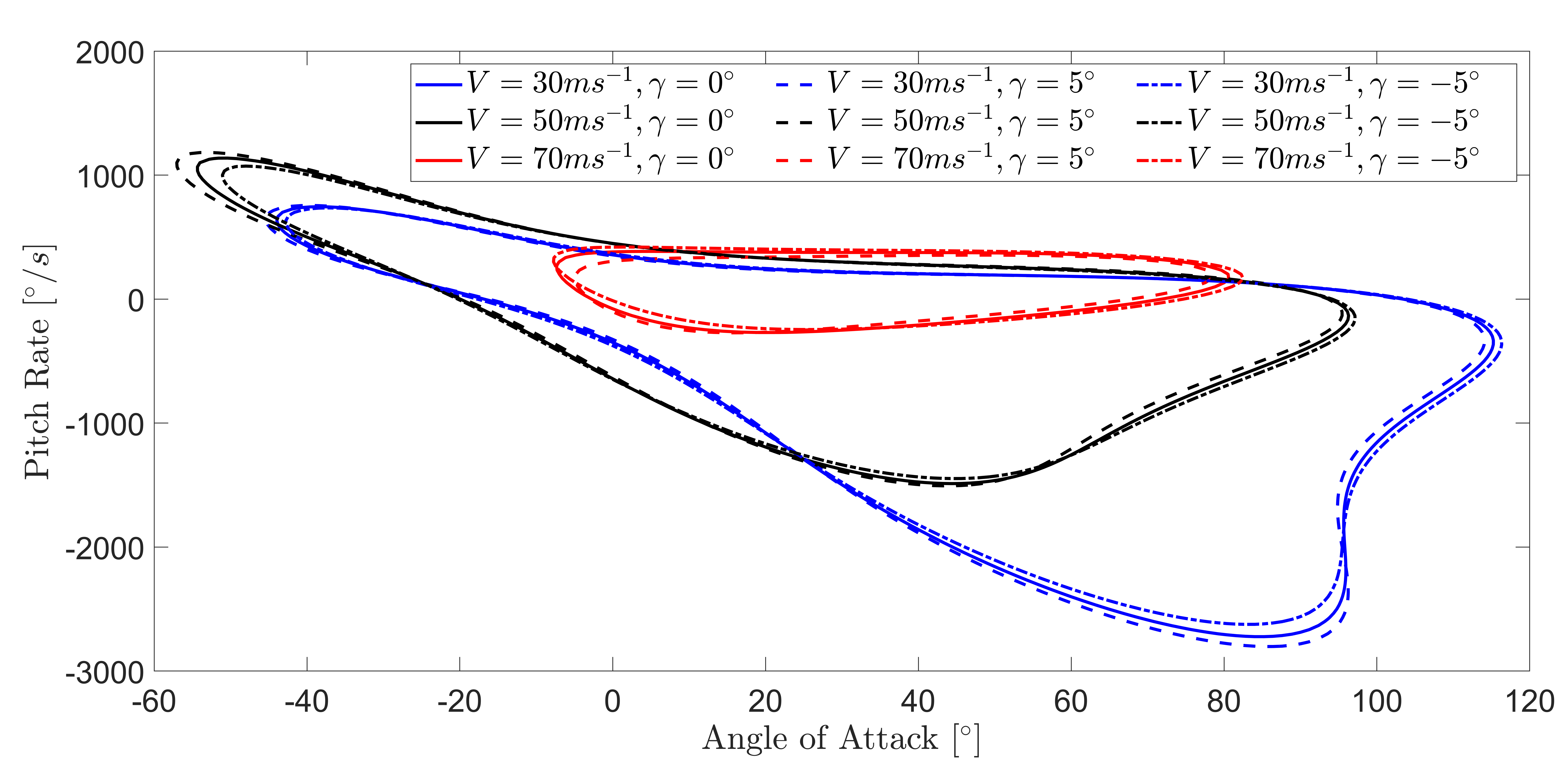}
		\caption{Level sets of 4D estimated NASA GTM longitudinal dynamics ROA. Level sets plotted onto $[a,q]$ space for fixed values of $[V_\mathrm{A}, \gamma_\mathrm{A}].$}  
		\label{GTMROA}               
\end{figure}

The ROA of this system is estimated from the 8565 trajectories using the algorithm presented above with monomials of degree 2 or less, i.e. a polynomial Lyapunov function of degree 4. Due to the large number of initial conditions, trajectory partitioning is not used, but constraints are implemented to avoid divergent ICs as detailed in equation (\ref{Semi Definite Prog Problem 2}). Figure \ref{GTMROA} shows several level sets of the four-state ROA. These level sets are for fixed values of $[V_\mathrm{A}, \gamma_\mathrm{A}]$ and plotted onto the $[a,q]$ space. The Lyapunov function obtained is given in Appendix section A. 

It is apparent there are convergent trajectories from very large angles of attack. This may be a limitation of studying longitudinal motion only and imposing extreme initial conditions of angle of attack and pitch rate. In particular, during wing stall at such extreme conditions, an aircraft rolls to either side, further destabilizing the aircraft \cite{gill}. Such effects are not captured in longitudinal motion and may explain the convergent trajectories from extremely positive angle of attack and extremely negative pitch rate, or vice versa. 

\begin{figure}[ht]
       \centering
		\includegraphics[width=0.6\textwidth]{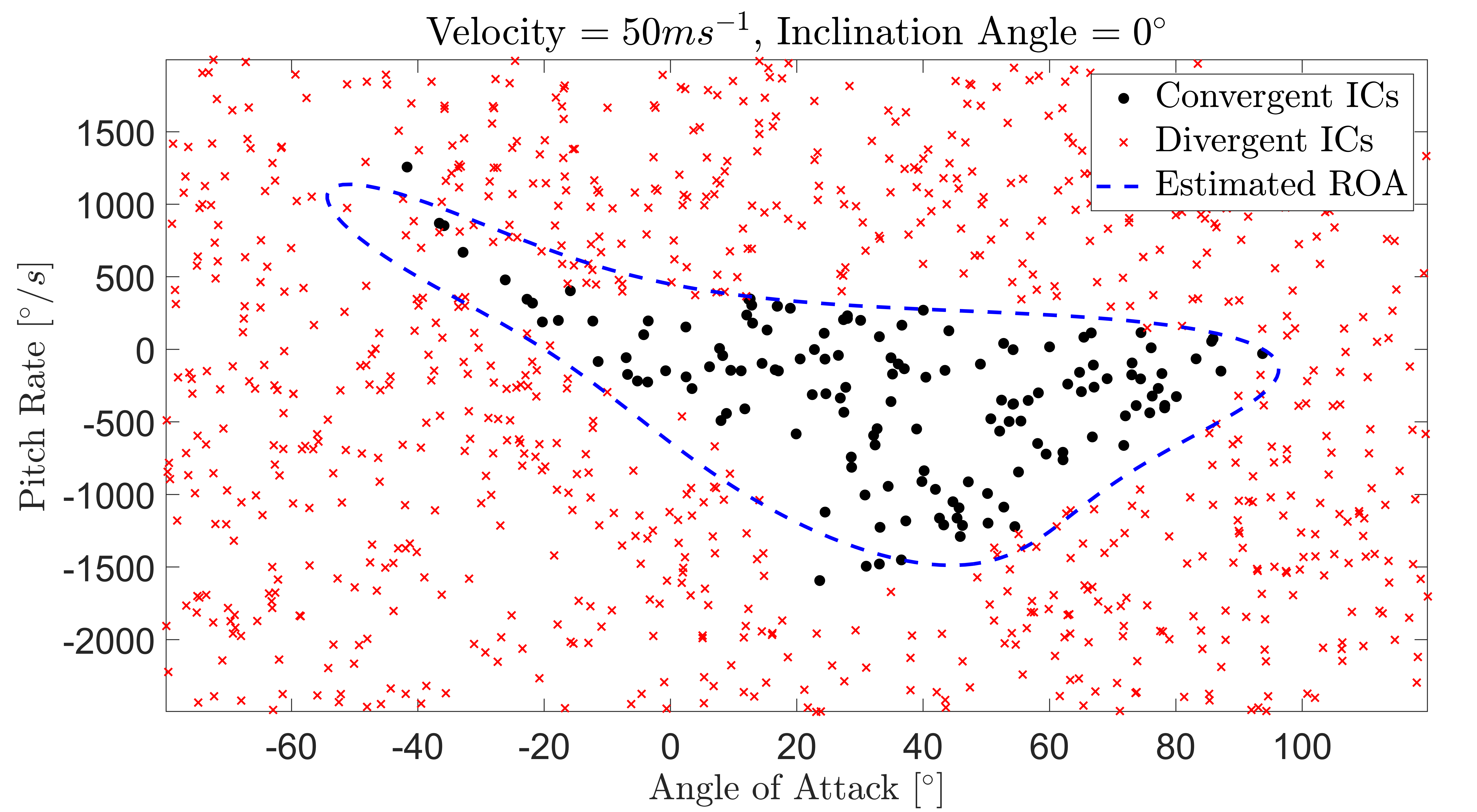}
		\caption{1000 additional ICs simulated with fixed inital $V_\mathrm{A} = \SI{50}{\meter\per\second}, \gamma_\mathrm{A} = \SI{0}{\degree}$ and variable $\alpha$ and $q$. ICs plotted against $V_\mathrm{A} = \SI{50}{\meter\per\second}, \gamma_\mathrm{A} = \SI{0}{\degree}$ level set of ROA estimation.}  
		\label{ROAValidation}               
\end{figure}

To validate the results, the level set defined by $V_\mathrm{A} = \SI{50}{\meter\per\second}, \gamma_\mathrm{A} = \SI{0}{\degree}$ is first considered. An additional 1000 ICs are tested, all starting from $V_\mathrm{A} = \SI{50}{\meter\per\second}, \gamma_\mathrm{A} = \SI{0}{\degree}$. These ICs are classified as convergent or divergent, and plotted against the estimated ROA shown in figure \ref{ROAValidation}. This process is repeated for the other eight levels sets shown in figure \ref{GTMROA}. These results are documented in appendix section C. 

\subsection{Convergence Analysis}

To perform the semi-definite optimization problem (\ref{Semi Definite Prog Problem 2}), the solver SeDuMi is used \cite{sedumi}. However, due to the large number of constraints in the problem (three constraints per trajectory), and size of matrix $P \in \mathbb{R}^{15 \times 15}$, a matrix P satisfying all constraints was not found. The following sections quantify how close the found $P^*$ is to satisfying all constraints and future steps to improve the solution. 

\subsubsection{Avoiding Divergent ICs Strict Constraint}

\begin{figure}[ht]
       \centering
		\includegraphics[width=\textwidth]{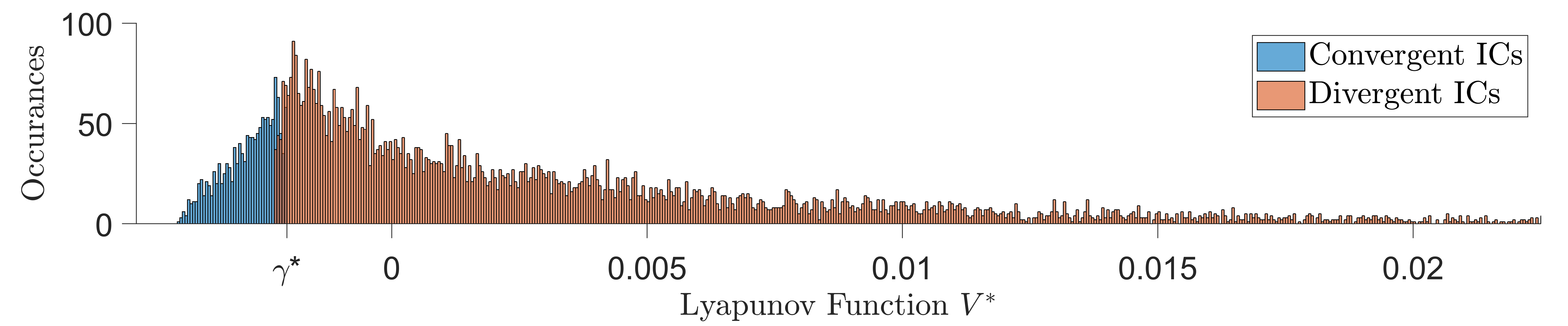}
		\caption{Histogram with bin size \SI{5e-5}{} of estimated Lyapunov Function $V^*$ evaluated for all known convergent and divergent initial conditions. The estimated ROA is $\{x \in \mathcal{X} \ : \ V^*(x) \leq \gamma^*\}$ where $\mathcal{X} = [V_\mathrm{A},\gamma_\mathrm{A},q,\alpha]^\mathrm{T}$. 6964 of the 7225 known divergent trajectories are successfully outside the estimated ROA.}  
		\label{Histogram}               
\end{figure}

Adding a constraint that all divergent ICs must be outside the estimated ROA in problem (\ref{Semi Definite Prog Problem 2}) in practice adds a strict constraint per divergent IC, $x_{d_i}$, that $Z_d(x_{d_i})^\mathrm{T} P Z_d(x_{d_i}) > \max\{Z_d(x_i)^\mathrm{T} P Z_d(x_i)\}$. Due to limitations of polynomial degree, this constraint is increasingly difficult or impossible to satisfy with a larger set of divergent ICs. Note that the estimate shown in figure \ref{GTMROA}, and in Appendix section A and C, are limited to a polynomial Lyapunov function of degree 4. Figure \ref{Histogram} shows the distribution of $V^*$ values for convergent and divergent ICs. 6964 of the 7225 known divergent trajectories are successfully outside the estimated ROA. Steps are being taken to modify this strict constraint into minimizing the number of known divergent ICs inside the estimated ROA. 

\subsubsection{Non-Semi-Definite Positive $P^*$}

To guarantee that the estimated Lyapunov function $V^*$ is SOS, a semi-definite positive matrix $P^*$ is required. Note that $P^*$ is SDP if and only if all eigenvalues of $P^*$ are non-negative. Appendix section B gives the eigenvalues of $P^*$, which shows $P^*$ is in fact not SDP, indicative of the negative $\gamma^*$ value. However, note that the smallest eigenvalue is \SI{-4.6221e-05}, thus relatively small in magnitude. Considering the closest SDP matrix in the Frobenius norm may be a suitable approximation.

\section{Extension to General ODE Systems}

Two additional algorithm extensions are presented to estimate the ROA of nonlinear ODE systems with more than a single equilibrium point. Examples such as the fall leaf mode show the importance of ROA estimation algorithms suitable for systems with general equilibrium sets. Although extensive linear and nonlinear analysis has been done on the falling leaf mode, works are limited to single equilibrium point analysis and may overlook nonlinear characteristics such as multiple equilibrium points and limit cycles \cite{Chakraborty_Falling_Leaf, Chakraborty_Falling_Leaf_nonlinear}.

\subsection{Finitely Many Equilibrium Points}
Consider a system of the form (\ref{ODE}), but now with a finite set of equilibrium points. That is, 
\begin{equation}
\begin{gathered}
    \label{ODE finitely many trim}
    \dot{x}(t) = f(x(t)), \ x(0) = x_0 \\
    f(x_{(i)}) = 0 \ : \ i = 1, ... ,l \ , \ l \in \mathbb{N}    
\end{gathered}
\end{equation}
The algorithm presented by Colbert is limited to finding the region of attraction to $x=0$, defined in (\ref{ROA 1 point}). There is further interest in estimating the region of attraction $S_{A_1}$ to a finite number of equilibrium conditions $A_1 \subseteq \{x_{(1)},...,x_{(l)}\}$ defined
\begin{equation}
\label{SA1}
    S_{A_1} = \{x \in \mathbb{R}^n \ : \ \exists x_{(i)} \in A_1 \ s.t. \ \lim_{t \to \infty}g(x,t) = x_{(i)} \}
\end{equation}

To find $S_{A_1}$, we are interested in finding a Lyapunov function that is zero exactly on $A_1$. That is, a continuously differentiable function $V_{A_1}:X \to \mathbb{R}^n$ on a compact set $X\subset \mathbb{R}^n$ with $A_1 \subset X$ such that

\begin{subequations}
\begin{align}
    V(x_{(i)}) = 0 & \quad \forall \ x_{(i)} \in A_1\\
    V(x) > 0 & \ \textnormal{for} \ x \in X, \ x \neq 0 \\
    \dot{V}(x) = \nabla V(x)^Tf(x) < 0 & \ \textnormal{for} \ x \in X, \ x \neq 0
\end{align}
\label{Lyapunov Theorem multi eq}
\end{subequations}

The goal of this algorithm is to approximate such a function (\ref{Lyapunov Theorem multi eq}) using the algorithm detailed in (\ref{Semi Definite Prog Problem}) and (\ref{Estimated Lyapunov Function}), and to estimate the ROA of (\ref{ODE finitely many trim}) using (\ref{Estimated Region of Attraction}).The algorithm improvements of trajectory partitioning and avoiding divergent ICs may similarly be implemented, but will be omitted from this section. 

Recall that to estimate Lyapunov function (\ref{Lyapunov Theorem}), a set of SOS polynomials of the form $p(x) = Z_d(x)^\mathrm{T} P Z_d(x)$ were considered, since $p(0) = 0$ and $p(x) > 0$. Next, trajectory data from a finite number of ICs is used to find SOS polynomial $p(x)$ most "similar" to the Lyapunov function form $V(x) = \int_{0}^{\infty}  \norm{g(x,t)}^2 dt$, i.e. of the form (\ref{Lyapunov Function}), via an SDP solver, detailed in (\ref{Semi Definite Prog Problem}). For an optimal $p(x)$, denoted $p^*(x)$, the property $\dot{p^*}(x) < 0$ is converged to since by construction, $\dot{V}(x) < 0$ for $V$ of the form (\ref{Lyapunov Function}). Finally, choosing Lyapunov function $V^*(x) = 10^{p^*(x)} - 1$ retains the properties of (\ref{Lyapunov Theorem multi eq}) while making numerical improvements detailed in \cite{Colbert}.

To extend to finitely many equilibrium points, the motivation is to first find a set of SOS polynomials $\{p_{A_1} | p_{A_1}(x) = 0 \ \forall x \in A_1\}$ to perform SDP optimization problem (\ref{Semi Definite Prog Problem}) within. To find a set of such polynomials, consider the following classical theorem and proof \cite{SOS_Theorem}. 
\begin{theorem}
Consider a polynomial $f(x_1,...,x_n)$ of degree $2d$. Let $z$ be a vector with all monomials of degree less than or equal to d. Then, $f$ is SOS whenever: 
$$f = z^\mathrm{T}Pz, \quad P \succeq 0$$
\end{theorem}
\noindent\textbf{Proof.} Since $P \succeq 0$, factor $P = L^\mathrm{T}L$. Then, 
$$f(x) = z^\mathrm{T}L^\mathrm{T}Lz = \norm{Lz}^2 = \sum_{i}(Lz)_i^2$$ \qed

Notice that for a SOS polynomial $f:\mathbb{R}^n \rightarrow \mathbb{R}$ where $f = z^\mathrm{T}Pz$, for any $x \in \mathbb{R}^n$, $f(x) = 0$ whenever $z_i(x) = 0$ for all $i$. Thus, to find the set of polynomials $\{p_{A_1} | p_{A_1}(x) = 0 \ \forall x \in A_1\}$, first find all monomials of degree $d$ or less that are zero exactly on $A_1$, detailed as follows: 
\par
Consider system (\ref{ODE finitely many trim}) in $\mathbb{R}^n$. Denote $x = [x_1,...,x_n]$ for any $x \in \mathbb{R}^n$, and each $x_{(i)} = [x_{1_{(i)}},...,x_{n_{(i)}}]$. Define the set 
\begin{equation}
  X_{(i)} = \{(x_1 - x_{1_{(i)}}),...,(x_n - x_{n_{(i)}}) \}
  \label{X_(i)}
\end{equation}
\noindent for each $x_{(i)} \in A_1$. Define $Z_{d,{A_1}}(x)$ as follows:
\begin{equation}
\begin{gathered}
    C_d(A_1) = \bigg\{\textnormal{Multisets} \ C \ \textnormal{comprised of elements of } \bigcup_{x_{(i)} \in A_1} X_{(i)} \ : \ |C| \leq d, \ C \cap X_{(i)} \neq \varnothing \quad \forall X_{(i)} \bigg\} \\
    Z_{d,{A_1}}(x) = \textnormal{The vector of all } \prod_{c \in C}c, \ \textnormal{where } C \in C_d(A_1) 
\end{gathered}
\label{Z_{d,{A_1}}}
\end{equation}

Note that from multiset theory, the length of vector $Z_{d,{A_1}}$ is $q_1 = \multiset{n|A_1|}{d} \coloneqq \binom{n|A_1|+d-1}{d}$. Notice that each entry of $Z_{d,{A_1}}(x)$ is $0$ exactly on $A_1$. Thus, for $P_{A_1} \succeq 0, \ Z_{d,{A_1}}(x)^\mathrm{T} P_{A_1} Z_{d,{A_1}}(x)$ is SOS and $0$ exactly on $A_1$. So, let $p_{A_1} = Z_{d,{A_1}}(x)^\mathrm{T} P_{A_1} Z_{d,{A_1}}(x)$. This set of polynomials is used in the same optimization detailed previously to find the ROA of (\ref{ODE finitely many trim}). In particular, the set of convergent initial conditions $\{x_i \ | \ i = 1,...,m\}$ is mapped to the set of outputs $\{y_i = \log_{10}{(1+V(x_i))} \ | \ i = 1,...,m\}$ for $V$ of the form (\ref{Lyapunov Function}) via SOS polynomial $Z_{d,{A_1}}(x)^\mathrm{T} P_{A_1} Z_{d,{A_1}}(x)$ in SDP optimization problem (\ref{Semi Definite Prog Problem multi eq}).


\begin{equation}
\label{Semi Definite Prog Problem multi eq}
\begin{split}
    \displaystyle\min_{P_{A_1}\in \mathbb{R}^{q_1\times q_1}, \ \gamma_i \in \mathbb{R}^m} \ & -\sum\limits_{i=1}^{m} \gamma_i \\
    \textnormal{s.t.} \quad & Z_{d,{A_1}}(x_i)^\mathrm{T} P_{A_1} Z_{d,{A_1}}(x_i) - y_i \geq \gamma_i \ \\
    & y_i - Z_{d,{A_1}}(x_i)^\mathrm{T} P_{A_1} Z_{d,{A_1}}(x_i) \geq \gamma_i \ ; \ P_{A_1}  \succeq 0.
\end{split}
\end{equation}

SeDuMi is used to find an optimal $P_{A_1}$, denoted $P_{A_1}^*$. Finally, the Lyapunov function of form (\ref{Lyapunov Function}) may be estimated by 
\begin{equation}
\label{Estimated Lyapunov Function multi eq}
    V_{A_1}^*(x) = 10^{Z_{d,{A_1}}(x_i)^\mathrm{T} P_{A_1} Z_{d,{A_1}}d(x_i)} - 1 
\end{equation}

Denoting $\gamma^*_1 = \displaystyle\max_{i , \in , 1,...,m}\{V_{A_1}^*(x_i)\}$, the estimated region of attraction $V_{\gamma^*,A_1}$ is expressed as

\begin{equation}
\label{Estimated Region of Attraction multi eq}
V_{\gamma^*,A_1} = {\{x \ : \ V_{A_1}^*(x)\leq \gamma^*_1\}}
\end{equation}

\begin{figure}[ht]
       \centering
		\includegraphics[width=0.5\textwidth]{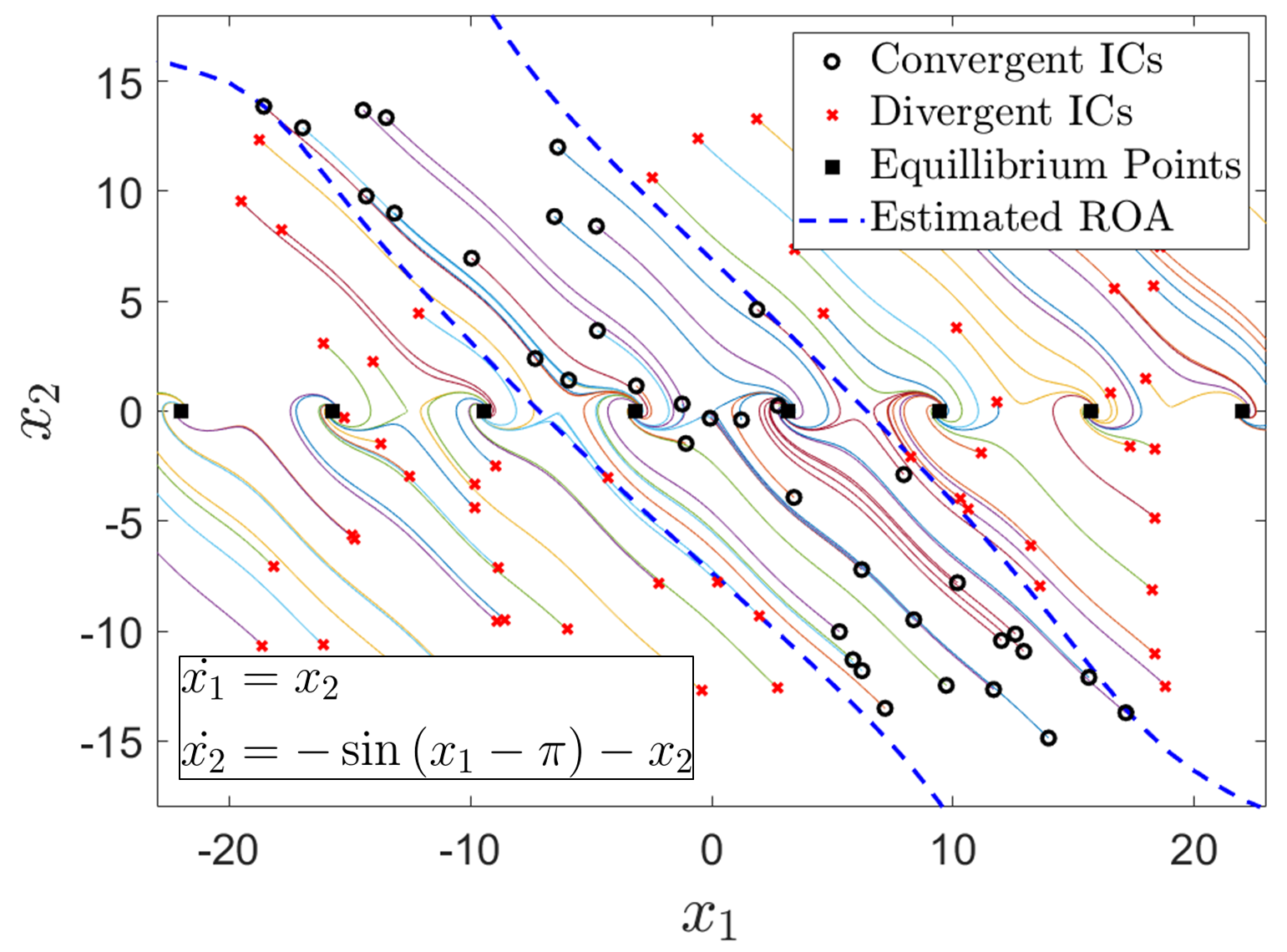}
		\caption{Estimated ROA to equillibrium points $(-\pi,0)$ and $(\pi,0)$ for the damped pendulum ODE.}  
		\label{pendulum}               
\end{figure}

This method has been used to estimate the ROA of an example ODE system to two of its equilibrium points, shown in figure \ref{pendulum}. This method is best for estimating the ROA to a small subset of equilibrium points, as problem complexity increases drastically with more equilibrium points.

\subsection{General Equilibrium Set and Limit Cycles}
Even further, consider systems of the form (\ref{ODE}), with an arbitrary set of equilibrium points, $E \subset \mathbb{R}^n $ and set of limit cycles $C = \{c_1, c_2, ..., c_l\}$. That is,  
\begin{equation}
\begin{gathered}
    \label{ODE arbitrary trim}
    \dot{x}(t) = f(x(t)), \ x(0) = x_0 \\
    E = \{x \ | \ f(x) = 0\} \ , \\
    C = \{\textrm{limit cycles } c \ : \ i = 1, ... ,l \ , \ l \in \mathbb{N} \}    
\end{gathered}
\end{equation}
The goal is to find the region of attraction $S_{A_2}$ to a subset of equilibrium points $\bar{E} \subseteq E$, or to a subset of limit cycles $\bar{C} \subseteq C$.That is,
\begin{equation}
\label{SA2}
    S_{A_2} = \{x \in \mathbb{R}^n \ : \  \exists \bar{x} \in \bar{E} \ s.t. \ \lim_{t \to \infty}g(x,t) = \bar{x} \quad \textnormal{or} \quad \exists \bar{c} \in \bar{C} \ s.t. \ \lim_{t \to \infty}g(x,t) \in \bar{c} \}    
\end{equation}
\begin{equation}
\label{A2}
    A_2 = \bar{E} \ \cup \left(\bigcup\limits_{\bar{c} \in \bar{C}} \bar{c}\right)
\end{equation}
where $\lim_{t \to \infty}g(x,t) \coloneqq \infty$ for divergent trajectories. Denote $A_2$ the ``equilibrium set" of (\ref{ODE arbitrary trim}). To estimate $S_{A_2}$, begin by considering the set of SOS polynomials

\begin{equation}
    \{p_{A_2} \ : \ p_{A_2}(x) = Z_{d,{\bar{A}_2}}^\mathrm{T}(x)P_{A_2}Z_{d,{\bar{A}_2}}(x), \ {\bar{A}_2}\subset A_2, \ |{\bar{A}_2}|<\infty \}
\end{equation}

where $Z_{d,{\bar{A}_2}}$ is of the form (\ref{Z_{d,{A_1}}}). Denote $q_2 = \multiset{n|{\bar{A}_2}|}{d}$ the length of vector $Z_{d,{\bar{A}_2}}$. Assume the set $A_2$ is either known or able to be approximated. Construct a function $r(x)$ that is $0$ on $A_2$ and positive elsewhere. For example, 
\begin{equation}
    r(x) = \left\{
        \begin{array}{ll}
            0 & \quad x \in A_2 \\
            \inf{\{ \ \norm{x-a}^2 \ : \ a \in A_2\}} & \quad x \notin A_2
        \end{array}
    \right.
\end{equation}

where $\norm{x-a}^2$ refers to the Euclidean norm. Further, assume such a function $r(x)$ is a continuously differentiable, lipschitz function. The motivation of this algorithm is to map the set of initial conditions $\{x_i \ : \ i = 1,...,m\}$ to the set of outputs $\bigg\{y_i = \nicefrac{\log_{10}\left(1 + V(x_i) \right)}{ r(x_i)} \ : \ i = 1,...m\bigg\}$ via an optimal $p_{A_2}$, denoted $p^*_{A_2}$. Then, $V^*(x) = 10^{p^*_{A_2}(x)r(x)} - 1$ is the estimated Lyapunov function. More specifically, solving the following SDP optimization problem, 

\begin{equation}
\label{Semi Definite Prog Problem general eq}
\begin{split}
    \displaystyle\min_{P_{A_2}\in \mathbb{R}^{q_2\times q_2}, \ \gamma_i \in \mathbb{R}^m} \ & -\sum\limits_{i=1}^{m} \gamma_i \\
    \textnormal{s.t.} \quad & Z_{d,{{\bar{A}_2}}}(x_i)^\mathrm{T} P_{A_2} Z_{d,{\bar{A}_2}}(x_i) - y_i \geq \gamma_i \ \\
    & y_i - Z_{d,{\bar{A}_2}}(x_i)^\mathrm{T} P_{A_2} Z_{d,{\bar{A}_2}}(x_i) \geq \gamma_i \ ; \ P_{A_2}  \succeq 0.
\end{split}
\end{equation}

gives an optimal $P_{A_2}$, denoted $P^*_{A_2}$. Notice, since $r(x)$ is constructed to be $0$ on $A_2$ and positive elsewhere, and since $p^*_{A_2}(x) = Z_{d,{\bar{A}_2}}^\mathrm{T}(x)P_{A_2}^*Z_{d,{\bar{A}_2}}(x)$ is $0$ on a subset of $A_2$ and positive elsewhere, then $r(x)p^*_{A_2}(x)$ is $0$ on $A_2$ and positive elsewhere. Thus, the Lyapunov function of form (\ref{Lyapunov Function}) may be estimated by

\begin{equation}
\label{Estimated Lyapunov Function general eq}
    V_{A_2}^*(x) = 10^{\big(Z_{d,{\bar{A}_2}}^\mathrm{T}(x)P_{A_2}^*Z_{d,{\bar{A}_2}}(x)\big)\big(r(x)\big)} - 1 
\end{equation}

Denoting $\gamma^*_2 = \displaystyle\max_{i , \in , 1,...,m}\{V_{A_2}^*(x_i)\}$, the estimated region of attraction $V_{\gamma^*,A_2}$ is expressed as

\begin{equation}
\label{Estimated Region of Attraction general eq}
V_{\gamma^*,A_2} = {\{x \ : \ V_{A_2}^*(x)\leq \gamma^*_2\}}
\end{equation}

An example system has not yet been tested with this proposed algorithm.

\section{Conclusion}
The major results of this work are improvements to the ROA estimation algorithm \cite{Colbert}, extension of this technique to ODEs of general equilibrium sets, and application to a 4D aircraft dynamics example. Future plans are to study the ability of an upset recovery strategy to return an aircraft to linear dynamics using this approach. A major limitation of this approach is SDP optimization failure in higher dimensions using conventional SDP optimization software. Given optimization difficulties in this 4D analysis, optimization difficulty is likely in higher dimensional problems using this approach. Future steps may be taken to increase robustness in more complex systems.

\newpage

\section*{Appendix}
\subsection{NASA GTM Longitudinal Dynamics ROA}
Equations (\ref{GTM Coefficients}) and (\ref{Gamma Star}) give the approximated polynomial degree 4 Lyapunov function and ROA estimate of the NASA GTM longitudinal dynamics. Recall from equations (\ref{Estimated Lyapunov Function}) and (\ref{Estimated Region of Attraction}) that the estimated Lyapunov function is given in the form $V^*(x) = 10^{Z_d(x)^T P^* Z_d(x)} - 1$   with the estimated ROA $V_{\gamma^*} = \{x \in \mathcal{X} \ | \ V^*(x) \leq \gamma^*\}$ where $\mathcal{X} = [V_\mathrm{A},\gamma_\mathrm{A},q,\alpha]^\mathrm{T}$.

\begin{equation}
\label{GTM Coefficients}
\begin{aligned}
Z_d(x)^\mathrm{T} P^* Z_d =&  \ \SI{5.2343e-10}{}V_\mathrm{A} ^4-\SI{6.9491e-09}{}V_\mathrm{A} ^3\gamma_\mathrm{A} -\SI{9.5519e-10}{}V_\mathrm{A} ^3q -\SI{7.9173e-09}{}V_\mathrm{A} ^3\alpha \\ &
     -\SI{8.3436e-08}{}V_\mathrm{A} ^3+\SI{5.5448e-07}{}V_\mathrm{A} ^2\gamma_\mathrm{A} ^2+\SI{7.7846e-08}{}V_\mathrm{A} ^2\gamma_\mathrm{A} q +\SI{4.8783e-07}{}V_\mathrm{A} ^2\gamma_\mathrm{A} \alpha \\ &
     +\SI{7.9070e-07}{}V_\mathrm{A} ^2\gamma_\mathrm{A} +\SI{6.9786e-10}{}V_\mathrm{A} ^2q ^2-\SI{3.9187e-08}{}V_\mathrm{A} ^2q \alpha +\SI{1.2171e-07}{}V_\mathrm{A} ^2q \\ &
     -\SI{7.2872e-08}{}V_\mathrm{A} ^2\alpha ^2+\SI{1.2863e-06}{}V_\mathrm{A} ^2\alpha +\SI{4.7225e-06}{}V_\mathrm{A} ^2+\SI{7.6022e-05}{}V_\mathrm{A} \gamma_\mathrm{A} ^3\\ &
     -\SI{1.2887e-06}{}V_\mathrm{A} \gamma_\mathrm{A} ^2q -\SI{1.2381e-05}{}V_\mathrm{A} \gamma_\mathrm{A} ^2\alpha -\SI{3.1231e-05}{}V_\mathrm{A} \gamma_\mathrm{A} ^2-\SI{6.4440e-08}{}V_\mathrm{A} \gamma_\mathrm{A} q ^2\\ &
     -\SI{1.1728e-06}{}V_\mathrm{A} \gamma_\mathrm{A} q \alpha -\SI{6.3809e-06}{}V_\mathrm{A} \gamma_\mathrm{A} q -\SI{3.5484e-06}{}V_\mathrm{A} \gamma_\mathrm{A} \alpha ^2-\SI{3.5022e-05}{}V_\mathrm{A} \gamma_\mathrm{A} \alpha \\ &
     -\SI{2.8717e-05}{}V_\mathrm{A} \gamma_\mathrm{A} -\SI{1.4972e-09}{}V_\mathrm{A} q ^3+\SI{1.6747e-08}{}V_\mathrm{A} q ^2\alpha -\SI{1.0734e-07}{}V_\mathrm{A} q ^2\\ &
     -\SI{8.3259e-07}{}V_\mathrm{A} q \alpha ^2+\SI{1.9745e-06}{}V_\mathrm{A} q \alpha -\SI{4.7537e-06}{}V_\mathrm{A} q +\SI{6.7208e-07}{}V_\mathrm{A} \alpha ^3\\ &
     +\SI{1.0927e-05}{}V_\mathrm{A} \alpha ^2-\SI{6.6253e-05}{}V_\mathrm{A} \alpha -\SI{1.1424e-04}{}V_\mathrm{A} +\SI{2.7521e-03}{}\gamma_\mathrm{A} ^4\\ &
     +\SI{4.7169e-05}{}\gamma_\mathrm{A} ^3q -\SI{8.3379e-04}{}\gamma_\mathrm{A} ^3\alpha -\SI{3.6243e-03}{}\gamma_\mathrm{A} ^3+\SI{1.0598e-06}{}\gamma_\mathrm{A} ^2q ^2\\ &
     -\SI{3.5378e-05}{}\gamma_\mathrm{A} ^2q \alpha +\SI{4.5128e-05}{}\gamma_\mathrm{A} ^2q -\SI{2.5809e-04}{}\gamma_\mathrm{A} ^2\alpha ^2+\SI{8.5833e-04}{}\gamma_\mathrm{A} ^2\alpha \\ &
     +\SI{3.2805e-04}{}\gamma_\mathrm{A} ^2+\SI{4.7299e-08}{}\gamma_\mathrm{A} q ^3-\SI{3.7614e-08}{}\gamma_\mathrm{A} q ^2\alpha +\SI{2.4142e-06}{}\gamma_\mathrm{A} q ^2\\ &
     -\SI{1.4365e-05}{}\gamma_\mathrm{A} q \alpha ^2+\SI{4.0195e-05}{}\gamma_\mathrm{A} q \alpha +\SI{1.1267e-04}{}\gamma_\mathrm{A} q +\SI{9.3334e-05}{}\gamma_\mathrm{A} \alpha ^3\\ &
     -\SI{6.1156e-05}{}\gamma_\mathrm{A} \alpha ^2+\SI{6.4515e-04}{}\gamma_\mathrm{A} \alpha +\SI{4.5015e-04}{}\gamma_\mathrm{A} +\SI{3.6012e-09}{}q ^4\\ &
     +\SI{1.6682e-07}{}q ^3\alpha +\SI{1.3707e-07}{}q ^3+\SI{2.6525e-06}{}q ^2\alpha ^2+\SI{8.3494e-07}{}q ^2\alpha \\ &
     +\SI{5.4175e-06}{}q ^2-\SI{4.5768e-07}{}q \alpha ^3+\SI{1.7422e-05}{}q \alpha ^2+\SI{7.1045e-05}{}q \alpha \\ &
     +\SI{5.4948e-05}{}q +\SI{2.0448e-04}{}\alpha ^4-\SI{4.0793e-04}{}\alpha ^3-\SI{3.8434e-05}{}\alpha ^2\\ &
     +\SI{7.6324e-04}{}\alpha
\end{aligned}
\end{equation}
\begin{equation}
    \label{Gamma Star}
    \gamma^* \ = \ \SI{-2.0533e-3}{}
\end{equation}

\subsection{Eigenvalues of $P^*$}
Recall from equation (\ref{Estimated Lyapunov Function}) that the estimated Lyapunov function of the NASA GTM longitudinal dynamics is given in the form $V^*(x) = 10^{Z_d(x)^\mathrm{T} P^* Z_d(x)} - 1$. Let $[\lambda_1, ..., \lambda_{15}]$ be the eigenvalues of $P^* \in \mathbb{R}^{15 \times 15}$.
\begin{equation}
[\lambda_1,   ... , \lambda_{15}] =
\begin{array}[t]{ccccc}
 \bigl[ \SI{-4.6221e-05}{} & \SI{-4.4984e-05}{} & \SI{-4.2299e-05}{} & \SI{1.1160e-05}{} & \SI{6.8337e-05}{} \\
 \SI{1.0706e-04}{} & \SI{1.4114e-04}{} & \SI{2.6110e-04}{} & \SI{3.7451e-04}{} & \SI{5.7837e-04}{} \\
 \SI{8.7243e-04}{} & \SI{2.7812e-03}{} & \SI{3.7985e-03}{} & \SI{1.6866e-02}{} & \num[retain-zero-exponent=true]{3.8170e+00} \bigr]
 \end{array} 
\end{equation}

\newpage

\subsection{ROA Validation Level Sets}

\begin{figure}[h!]
\begin{subfigure}{.49\textwidth}
  \centering
  \includegraphics[width=\linewidth]{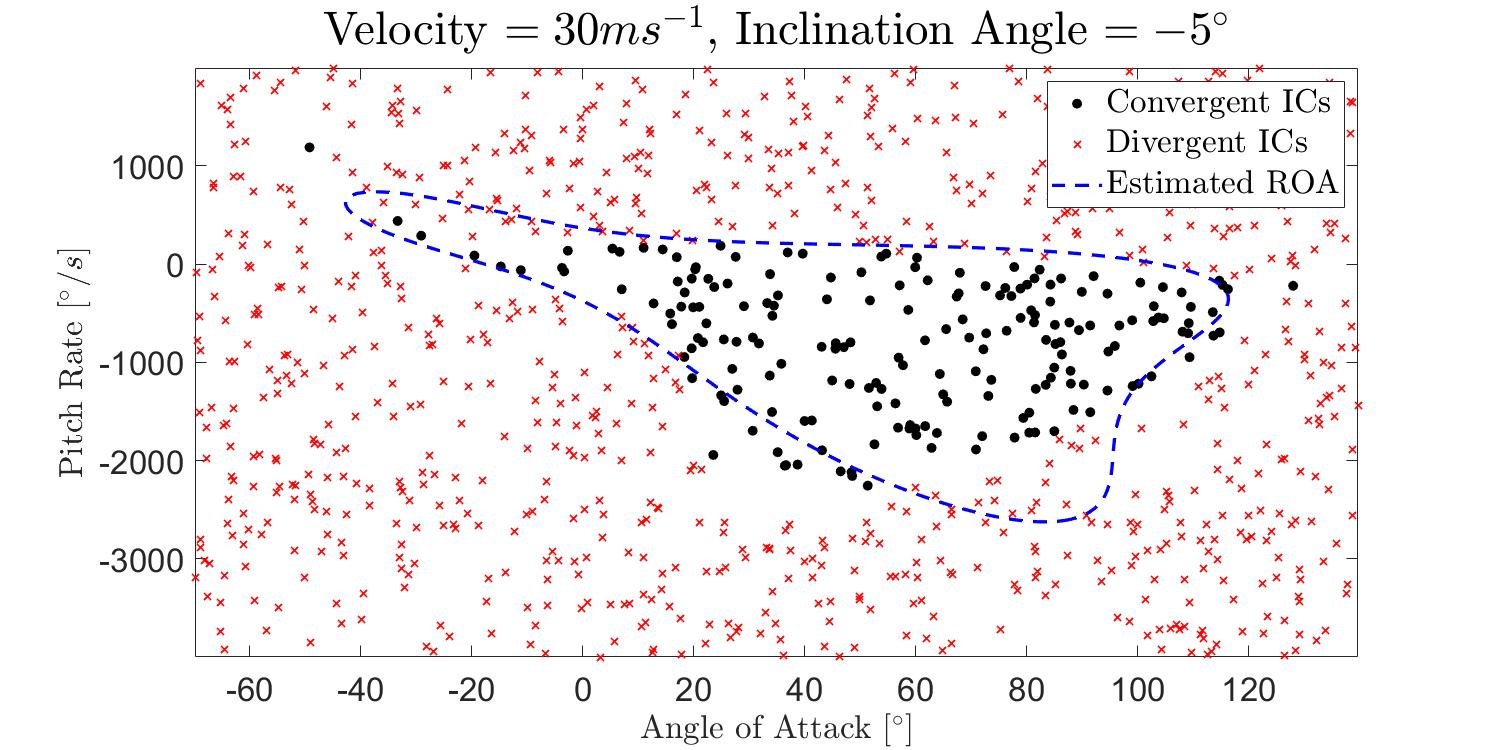}  
\end{subfigure}
\vspace{10pt}
\begin{subfigure}{.5\textwidth}
  \centering
  \includegraphics[width=\linewidth]{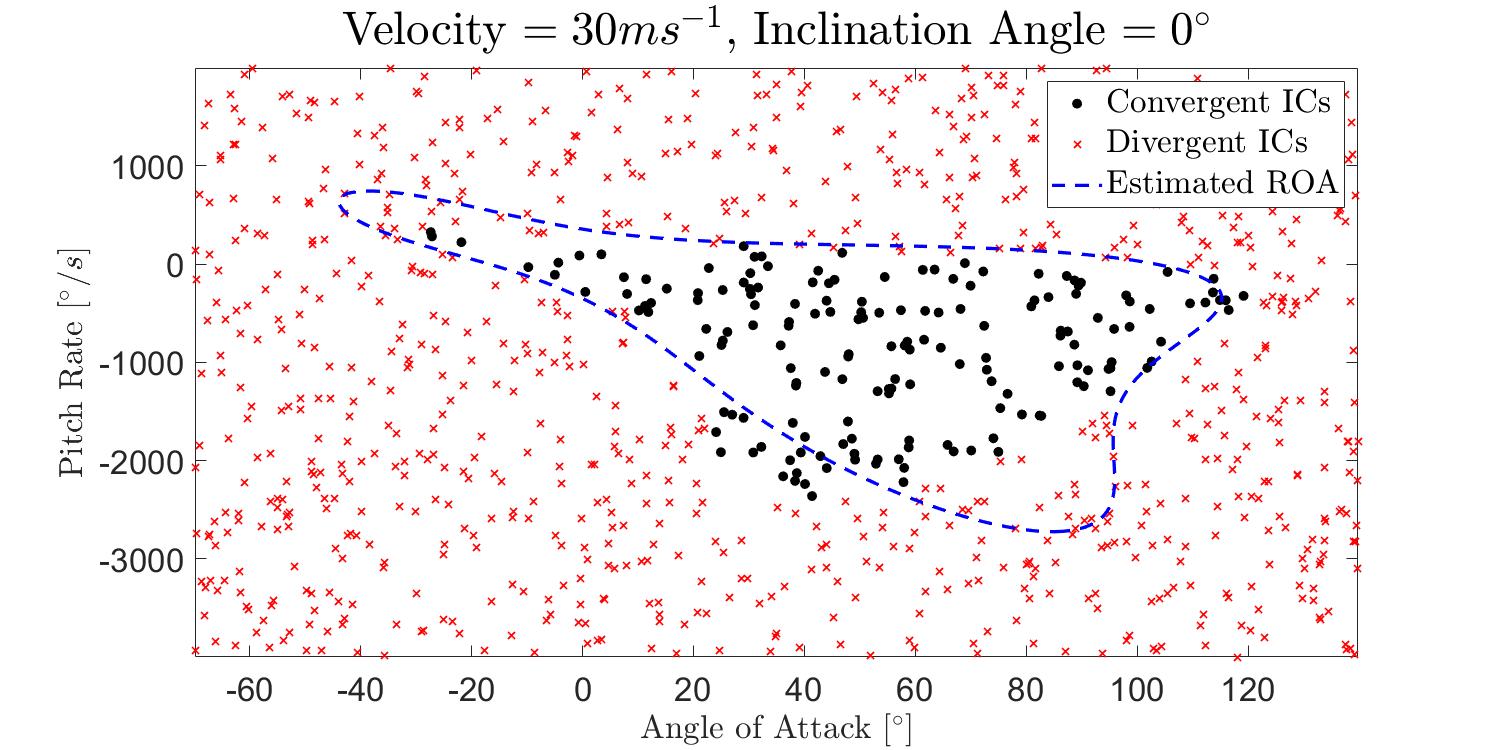}  
\end{subfigure}
\vspace{10pt}
\begin{subfigure}{.5\textwidth}
  \centering
  \includegraphics[width=\linewidth]{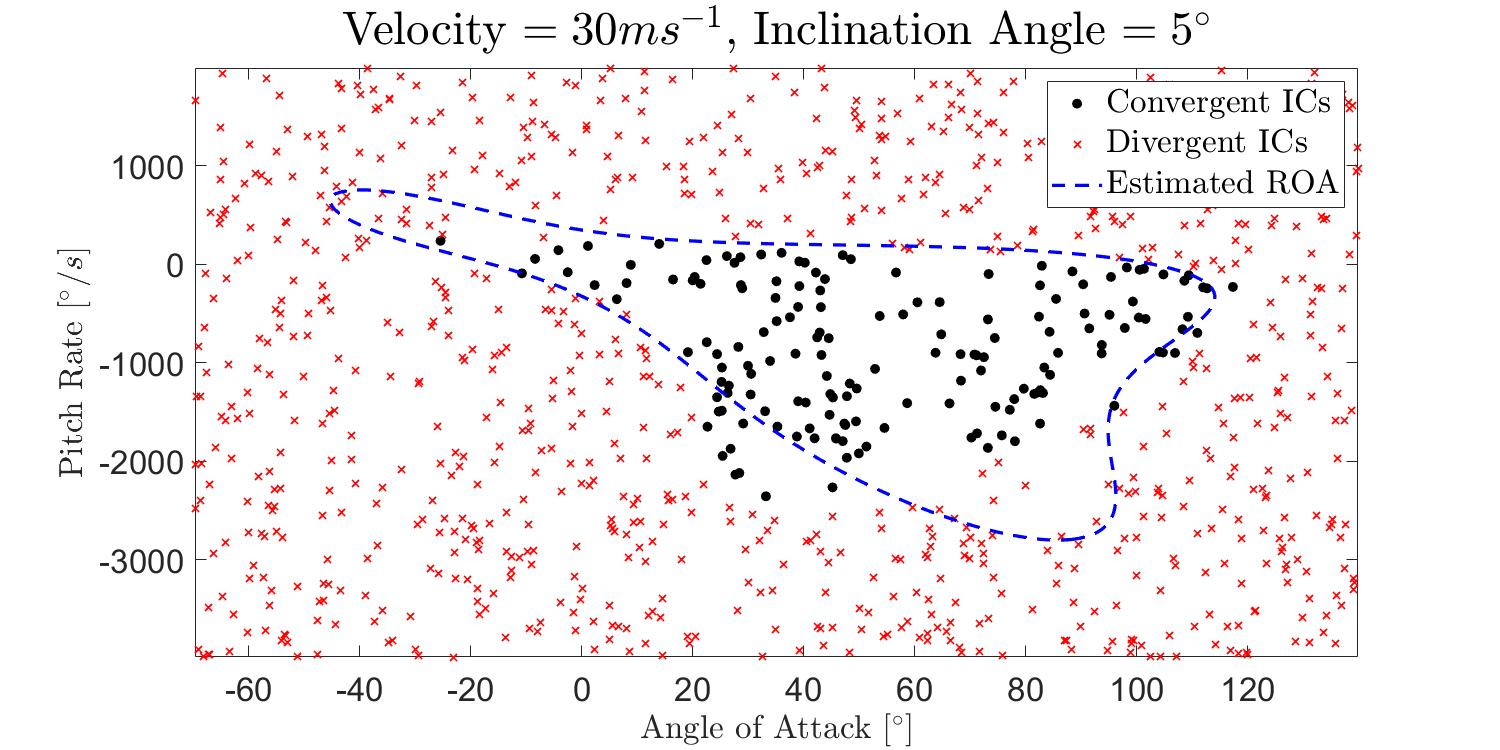}  
\end{subfigure}
\begin{subfigure}{.5\textwidth}
  \centering
  \includegraphics[width=\linewidth]{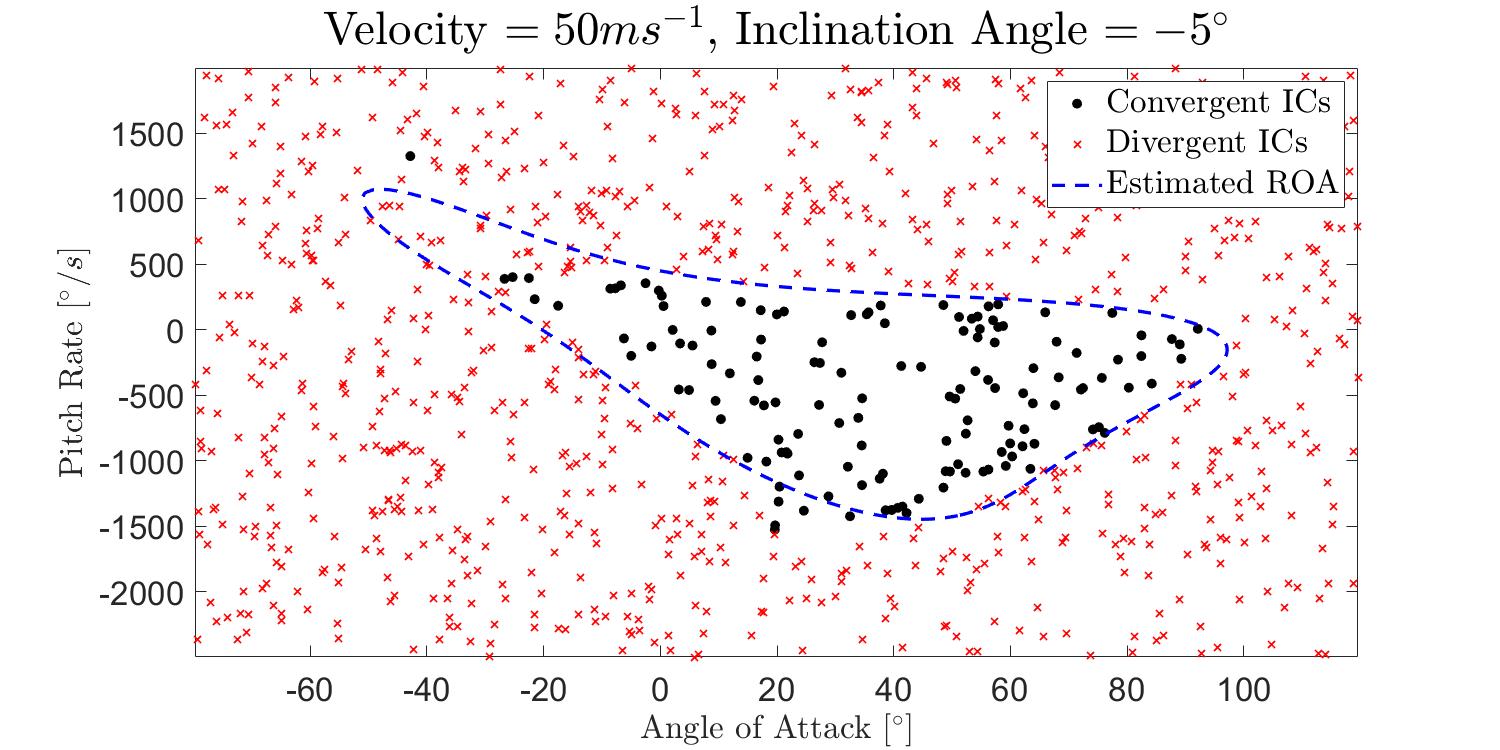}  
\end{subfigure}
\vspace{10pt}
\begin{subfigure}{.5\textwidth}
  \centering
  \includegraphics[width=\linewidth]{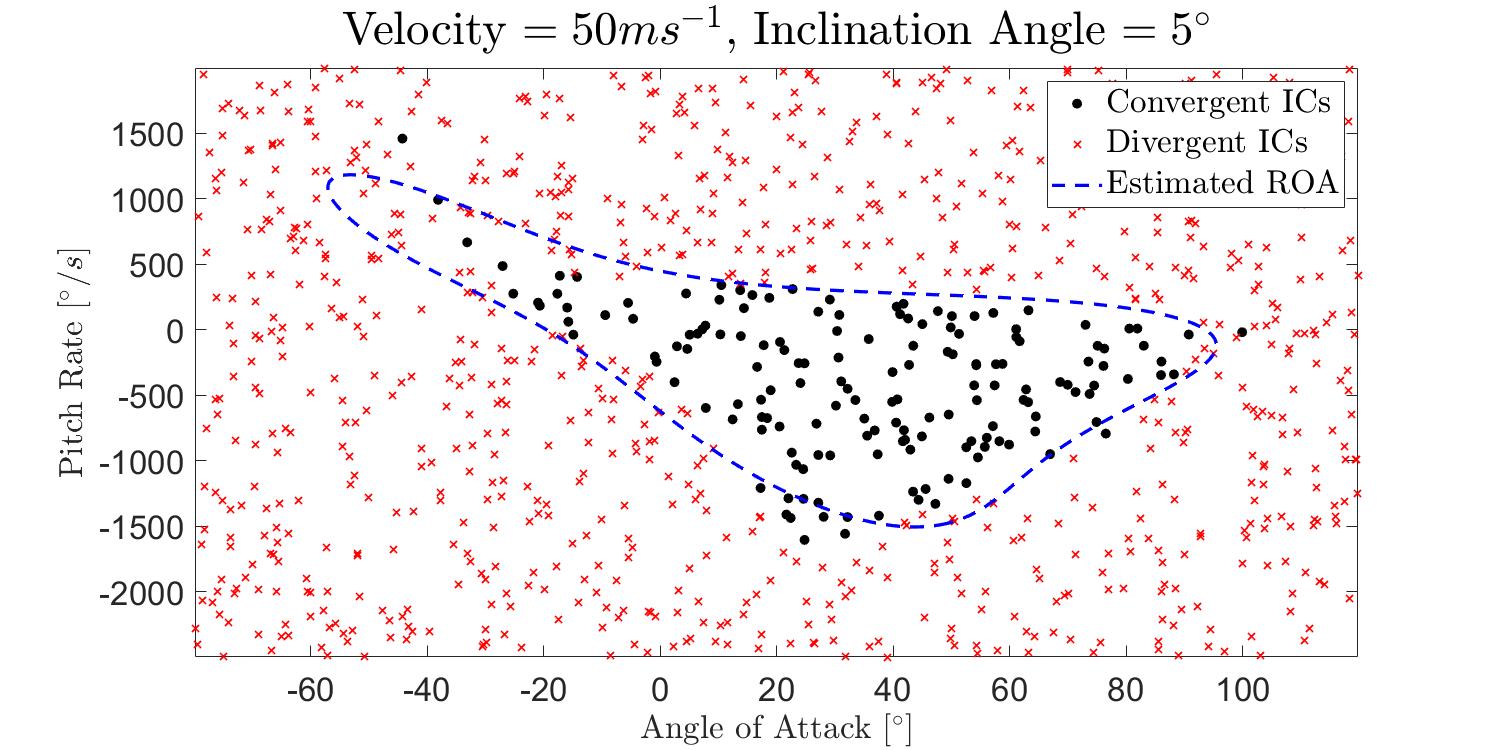}  
\end{subfigure}
\begin{subfigure}{.5\textwidth}
  \centering
  \includegraphics[width=\linewidth]{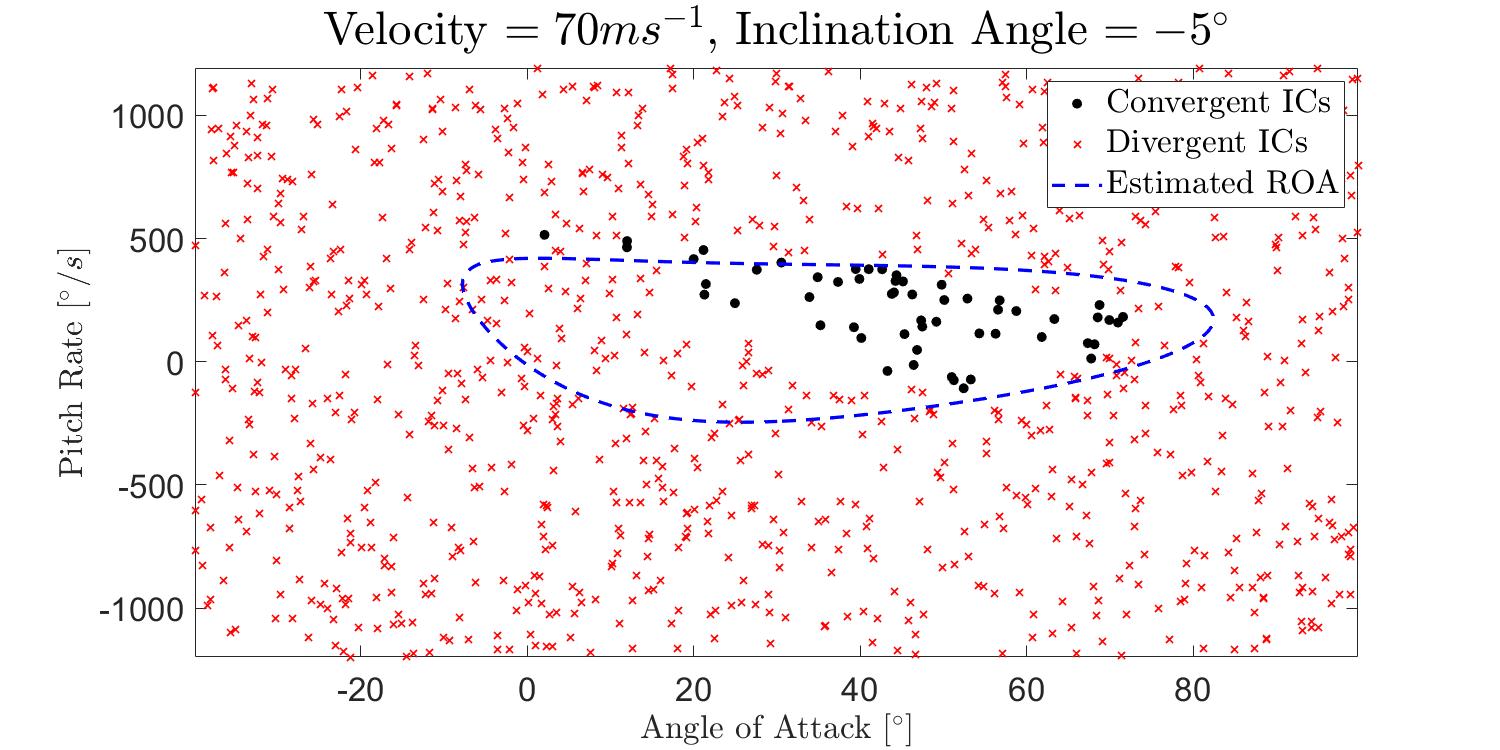}  
\end{subfigure}
\begin{subfigure}{.5\textwidth}
  \centering
  \includegraphics[width=\linewidth]{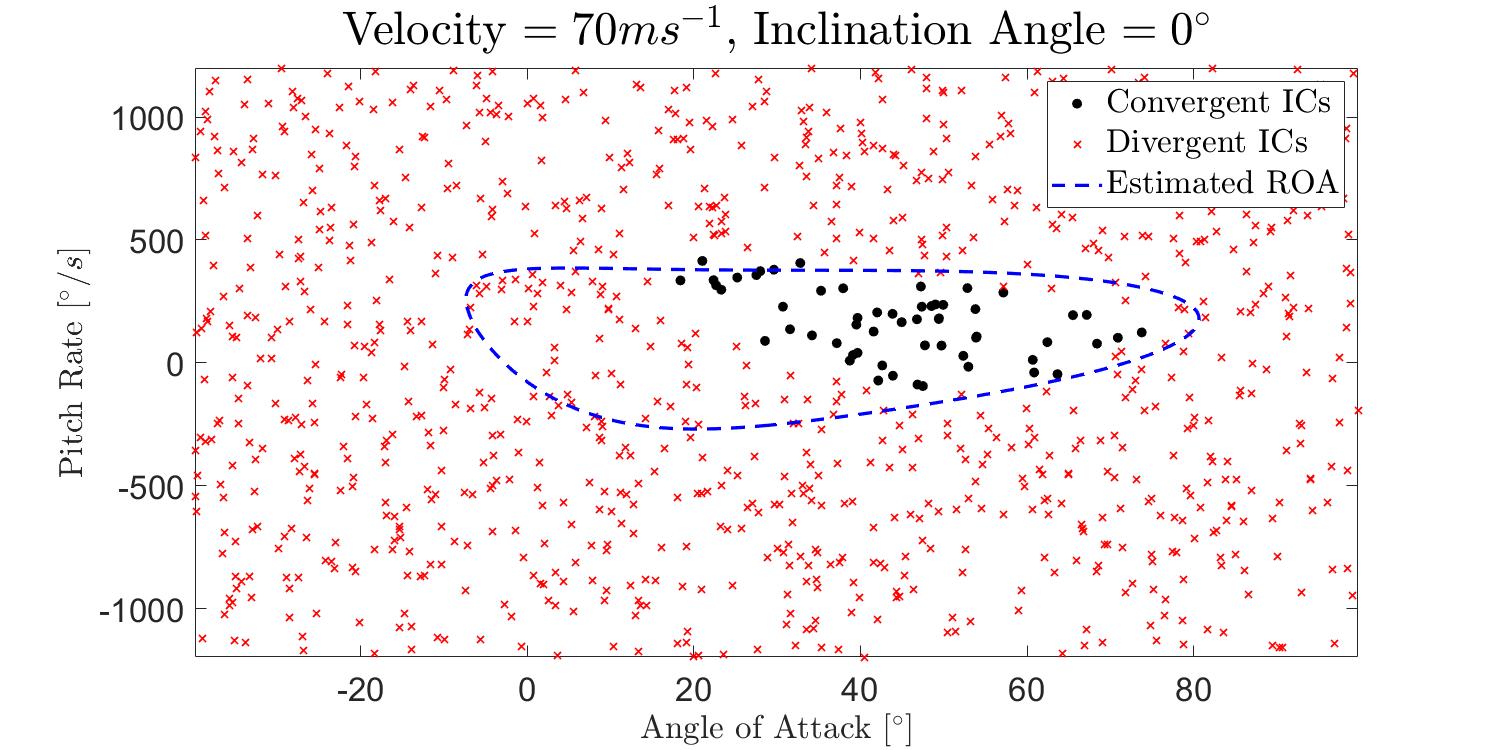}  
\end{subfigure}
\begin{subfigure}{.5\textwidth}
  \centering
  \includegraphics[width=\linewidth]{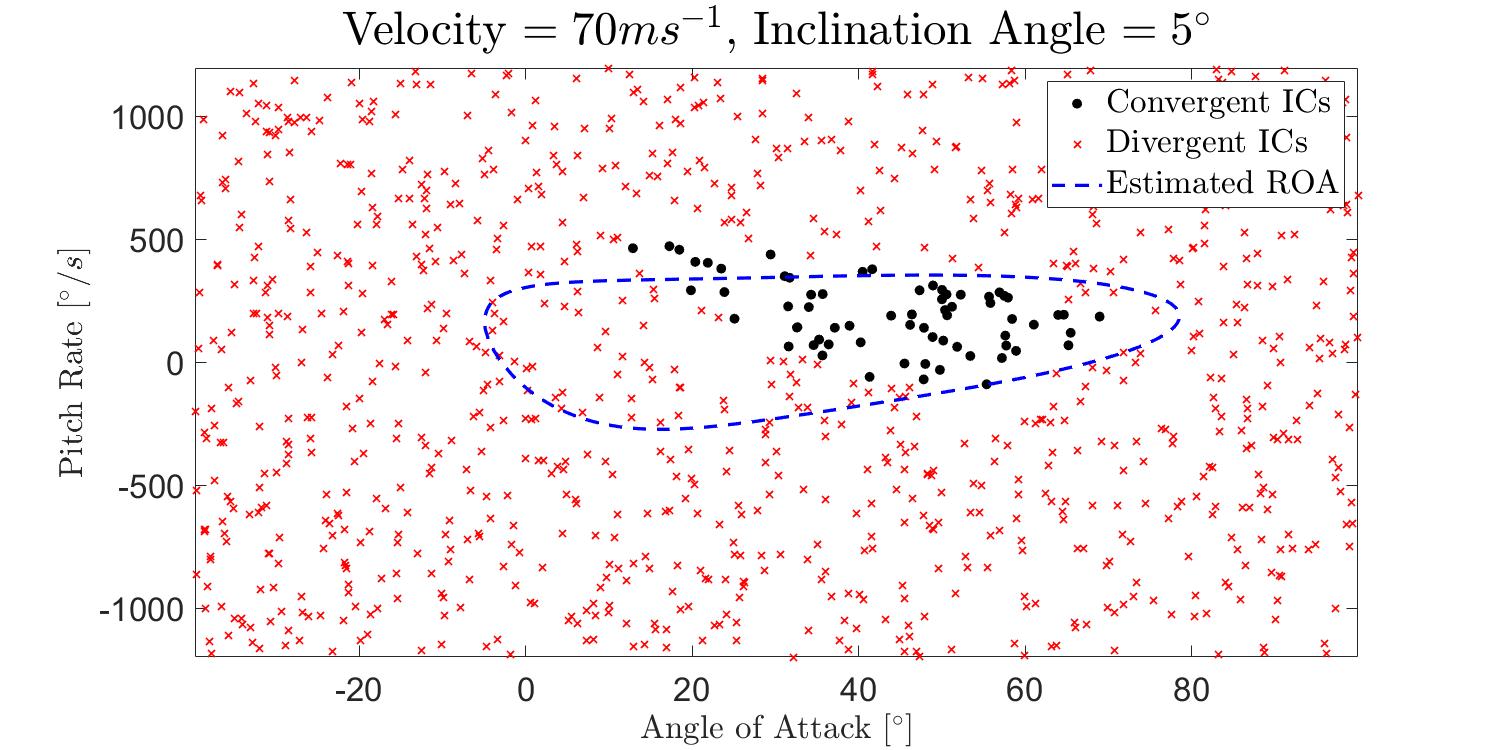}  
\end{subfigure}
\caption{1000 additional ICs simulated with fixed inital velocity and inclination angle labeled respectively and variable angle of attack and pitch rate. ICs plotted against the respective $(V_\mathrm{A},\gamma_\mathrm{A})$ level set of ROA estimation.}
\label{ROA Validation Appendix}
\end{figure}

\newpage

\section*{Acknowledgments}
This work was funded by the New Jersey NASA Space Grant Consortium. Additional support was provided by the Rutgers James J. Slade Scholars Program.

\bibliography{main}

\begin{thebibliography}{18}
\newcommand{\enquote}[1]{``#1''}
\providecommand{\natexlab}[1]{#1}
\providecommand{\url}[1]{\texttt{#1}}
\providecommand{\urlprefix}{URL }
\expandafter\ifx\csname urlstyle\endcsname\relax
  \providecommand{\doi}[1]{\discretionary{}{}{}https://doi.org/#1}\else
  \providecommand{\doi}[1]{\discretionary{}{}{}\urlstyle{rm}\url{https://doi.org/#1}}\fi

\bibitem[{Chakraborty et~al.(2011{\natexlab{a}})Chakraborty, Seiler, and
  Balas}]{Chakraborty_GTM}
Chakraborty, A., Seiler, P., and Balas, G., \enquote{Nonlinear region of
  attraction analysis for flight control verification and validation,}
  \emph{Control Engineering Practice}, Vol.~19, 2011{\natexlab{a}}, pp.
  335--345.
\newblock \doi{10.1016/j.conengprac.2010.12.001}.

\bibitem[{Goman et~al.(1997)Goman, Zagainov, and Khramtsovsky}]{Goman}
Goman, M., Zagainov, G., and Khramtsovsky, A., \enquote{Application of
  bifurcation methods to nonlinear flight dynamics problems,} \emph{Progress in
  Aerospace Sciences}, Vol.~33, 1997, pp. 539--586.
\newblock \doi{10.1016/S0376-0421(97)00001-8}.

\bibitem[{Chakraborty et~al.(2011{\natexlab{b}})Chakraborty, Seiler, and
  Balas}]{Chakraborty_Falling_Leaf}
Chakraborty, A., Seiler, P., and Balas, G., \enquote{Susceptibility of F/A-18
  Flight Controllers to the Falling-Leaf Mode: Linear Analysis,} \emph{Journal
  of Guidance, Control, and Dynamics}, Vol.~34, 2011{\natexlab{b}}, pp. 73--85.
\newblock \doi{10.2514/1.50675}.

\bibitem[{Chakraborty et~al.(2011{\natexlab{c}})Chakraborty, Seiler, and
  Balas}]{Chakraborty_Falling_Leaf_nonlinear}
Chakraborty, A., Seiler, P., and Balas, G.~J., \enquote{Susceptibility of
  F/A-18 Flight Controllers to the Falling-Leaf Mode: Nonlinear Analysis,}
  \emph{Journal of Guidance, Control, and Dynamics}, Vol.~34, No.~1,
  2011{\natexlab{c}}, pp. 73--85.
\newblock \doi{10.2514/1.50675},
  \urlprefix\url{https://doi.org/10.2514/1.50675}.

\bibitem[{Cunis et~al.(2018{\natexlab{a}})Cunis, Burlion, and
  Condomines}]{Cunis2019}
Cunis, T., Burlion, L., and Condomines, J.-P., \enquote{Piecewise Polynomial
  Modeling for Control and Analysis of Aircraft Dynamics Beyond Stall,}
  \emph{Journal of Guidance, Control, and Dynamics}, 2018{\natexlab{a}}.
\newblock \doi{10.2514/1.G003618}.

\bibitem[{Engelbrecht et~al.(2013)Engelbrecht, Pauck, and Peddle}]{GTM_ex1}
Engelbrecht, J.~A., Pauck, S.~J., and Peddle, I.~K., \enquote{A Multi-Mode
  Upset Recovery Flight Control System for Large Transport Aircraft,}
  \emph{AIAA Guidance, Navigation, and Control (GNC) Conference}, 2013.
\newblock \doi{10.2514/6.2013-5172},
  \urlprefix\url{https://arc.aiaa.org/doi/abs/10.2514/6.2013-5172}.

\bibitem[{Topcu et~al.(2010)Topcu, Packard, Seiler, and Balas}]{Topcu1}
Topcu, U., Packard, A., Seiler, P., and Balas, G., \enquote{Robust
  Region-of-Attraction Estimation,} \emph{Automatic Control, IEEE Transactions
  on}, Vol.~55, 2010, pp. 137 -- 142.
\newblock \doi{10.1109/TAC.2009.2033751}.

\bibitem[{Topcu and Packard(2008)}]{Topcu2}
Topcu, U., and Packard, A., \enquote{Stability region analysis for uncertain
  nonlinear systems,} \emph{Proceedings of the IEEE Conference on Decision and
  Control}, 2008, pp. 1693 -- 1698.
\newblock \doi{10.1109/CDC.2007.4434914}.

\bibitem[{Tan and Packard(2008)}]{Tan}
Tan, W., and Packard, A., \enquote{Stability Region Analysis Using Polynomial
  and Composite Polynomial Lyapunov Functions and Sum-of-Squares Programming,}
  \emph{Automatic Control, IEEE Transactions on}, Vol.~53, 2008, pp. 565 --
  571.
\newblock \doi{10.1109/TAC.2007.914221}.

\bibitem[{Parrilo(2000)}]{Parrilo}
Parrilo, P., \enquote{Structured Semidenite Programs and Semialgebraic Geometry
  Methods in Robustness and Optimization,} \emph{PhD thesis}, 2000.

\bibitem[{{Colbert} and {Peet}(2018)}]{Colbert}
{Colbert}, B.~K., and {Peet}, M.~M., \enquote{Using Trajectory Measurements to
  Estimate the Region of Attraction of Nonlinear Systems,} \emph{2018 IEEE
  Conference on Decision and Control (CDC)}, 2018, pp. 2341--2347.

\bibitem[{Topcu et~al.(2008)Topcu, Packard, and Seiler}]{Topcu2008}
Topcu, U., Packard, A., and Seiler, P., \enquote{Local stability analysis using
  simulations and sum-of-squares programming,} \emph{Automatica}, Vol.~44,
  2008, pp. 2669--2675.
\newblock \doi{10.1016/j.automatica.2008.03.010}.

\bibitem[{Cunis et~al.(2018{\natexlab{b}})Cunis, Burlion, and
  Condomines}]{GTM_piecewise}
Cunis, T., Burlion, L., and Condomines, J.-P., \enquote{{Piecewise Polynomial
  Model of the Aerodynamic Coefficients of the Generic Transport Model and its
  Equations of Motion},} Technical Report Third corrected version, {ONERA --
  The French Aerospace Lab ; {\'E}cole Nationale de l'Aviation Civile},
  2018{\natexlab{b}}.
\newblock \urlprefix\url{https://hal.archives-ouvertes.fr/hal-01808649}.

\bibitem[{Cox(2019)}]{NASA_GTM}
Cox, D., \enquote{The GTM DesignSim,} , 08 2019.
\newblock \urlprefix\url{https://github.com/nasa/GTM_DesignSim}.

\bibitem[{Sturm(1998)}]{sedumi}
Sturm, J., \enquote{Using SeDuMi 1.02, a MATLAB toolbox for optimization over
  symmetric cones,} \emph{Optimization Methods and Software}, Vol.~11, 1998.
\newblock \doi{10.1080/10556789908805766}.

\bibitem[{Sturm and Guerra(1999)}]{Sturm1999AMT}
Sturm, J.~F., and Guerra, G.~A., \enquote{A Matlab toolbox for optimization
  over symmetric cones,} , 1999.

\bibitem[{Gill et~al.(2013)Gill, Lowenberg, Neild, Krauskopf, Puyou, and
  Coetzee}]{gill}
Gill, S., Lowenberg, M., Neild, S., Krauskopf, B., Puyou, G., and Coetzee, E.,
  \enquote{Upset Dynamics of an Airliner Model: A Nonlinear Bifurcation
  Analysis,} \emph{Journal of Aircraft}, Vol.~50, No.~6, 2013, pp. 1832--1842.
\newblock \doi{10.2514/1.C032221}.

\bibitem[{{Seiler} and {Balas}(2010)}]{SOS_Theorem}
{Seiler}, P., and {Balas}, G.~J., \enquote{Quasiconvex sum-of-squares
  programming,} \emph{49th IEEE Conference on Decision and Control (CDC)},
  2010, pp. 3337--3342.

\end{thebibliography}

\end{document}